\documentclass[preprint]{aastex}
\usepackage[usenames]{color}

\begin{document}

\shortauthors{Welty et al.}
\shorttitle{M82 ISM:  DIBs vs. Atomic and Molecular Species}

%\slugcomment{DRAFT -- 2014 jul15h}

\title{Diffuse Interstellar Bands vs. Known Atomic and Molecular Species in the Interstellar Medium of M82 toward SN 2014J}

\author{Daniel E. Welty\altaffilmark{1}, Adam M. Ritchey\altaffilmark{2}, Julie A. Dahlstrom\altaffilmark{3}, Donald G. York\altaffilmark{1,4}}

\altaffiltext{1}{University of Chicago, Department of Astronomy and Astrophysics, 5640 S. Ellis Ave., Chicago, IL 60637; dwelty@oddjob.uchicago.edu}
\altaffiltext{2}{University of Washington, Department of Astronomy, Box 351580, Seattle, WA 98195}
\altaffiltext{3}{Carthage College, Department of Physics and Astronomy, 2001 Alford Park Dr., Kenosha, WI 53140}
\altaffiltext{4}{also, Enrico Fermi Institute}

\begin{abstract}

We discuss the absorption due to various constituents of the interstellar medium of M82 seen in moderately high resolution, high signal-to-noise ratio optical spectra of SN 2014J.
Complex absorption from M82 is seen, at velocities 45 $\la$ $v_{\rm LSR}$ $\la$ 260 km~s$^{-1}$, for \ion{Na}{1}, \ion{K}{1}, \ion{Ca}{1}, \ion{Ca}{2}, CH, CH$^+$, and CN; many of the diffuse interstellar bands (DIBs) are also detected.
Comparisons of the column densities of the atomic and molecular species and the equivalent widths of the DIBs reveal both similarities and differences in relative abundances, compared to trends seen in the ISM of our Galaxy and the Magellanic Clouds.
Of the ten relatively strong DIBs considered here, six (including $\lambda$5780.5) have strengths within $\pm$20\% of the mean values seen in the local Galactic ISM, for comparable $N$(\ion{K}{1}); two are weaker by 20--45\% and two (including $\lambda$5797.1) are stronger by 25--40\%.
Weaker than ''expected'' DIBs [relative to $N$(\ion{K}{1}), $N$(\ion{Na}{1}), and $E(B-V)$] in some Galactic sight lines and toward several other extragalactic supernovae appear to be associated with strong CN absorption and/or significant molecular fractions.
While the $N$(CH)/$N$(\ion{K}{1}) and $N$(CN)/$N$(CH) ratios seen toward SN 2014J are similar to those found in the local Galactic ISM, the combination of high $N$(CH$^+$)/$N$(CH) and high $W$(5797.1)/$W$(5780.5) ratios has not been seen elsewhere.
The centroids of many of the M82 DIBs are shifted, relative to the envelope of the \ion{K}{1} profile -- likely due to component-to-component variations in $W$(DIB)/$N$(\ion{K}{1}) that may reflect the molecular content of the individual components.
We compare estimates for the host galaxy reddening $E(B-V)$ and visual extinction $A_{\rm V}$ derived from the various interstellar species with the values estimated from optical and near-IR photometry of SN 2014J.

\end{abstract}


\section{INTRODUCTION}
\label{sec-intro}

Supernovae in external galaxies provide rare, fleeting opportunities to probe the interstellar media of the host galaxies via absorption-line spectroscopy.
In principle, such observations can reveal the behavior of various tracers of the ISM under somewhat different environmental conditions from those typically sampled in the local Galactic ISM.
Differences in overall metallicity, specific elemental abundance ratios, dust-to-gas ratios, and radiation fields all can affect the structure and composition of interstellar clouds (e.g., Wolfire et al. 1995; Pak et al. 1998); in combination, the overall effects can be somewhat unexpected and counter-intuitive.
Exploration of diverse environments, where those factors act in different combinations, thus can aid in disentangling the specific effects of each factor.

It is particularly important to compare the behavior of the enigmatic diffuse interstellar bands (DIBs) in a variety of environments with the corresponding behavior of known atomic and molecular constituents of the ISM.
Such comparisons should aid both in identifying the carriers of the DIBs and in calibrating the DIBs as diagnostics of the physical conditions in the ISM.
Some of the typically strongest DIBs (in the local Galactic ISM) have therefore been measured toward a small number of stars in the Large and Small Magellanic Clouds (LMC and SMC; Vladilo et al. 1987; Ehrenfreund et al. 2002; Cox et al. 2006, 2007; Welty et al. 2006), toward a much larger set of stars in the 30 Dor region of the LMC (van Loon et al. 2013), and toward some stars in M33 (Cordiner et al. 2008b) and M31 (Cordiner et al. 2008a, 2011).
A few DIBs have been detected in the host galaxies of extragalactic supernovae (D'Odorico et al. 1989; Sollerman et al. 2005; Cox \& Patat 2008, 2014) and in several damped Lyman-$\alpha$ systems (York et al. 2006; Ellison et al. 2008; Lawton et al. 2008); Phillips et al. (2013) have compiled measurements of the $\lambda$5780.5 DIB toward a number of recent Type Ia SNe.
Examination of the behavior of the DIBs in the Local Group galaxies suggests that the DIB strengths can depend both on the overall metallicity and on local physical conditions (e.g., radiation field, molecular fraction).
Certain DIB ratios (e.g., $\lambda$5797.1/$\lambda$5780.5) may provide information on the ambient radiation fields (e.g., Vos et al. 2011); the strengths of individual DIBs (e.g., $\lambda$5780.5) may yield useful estimates for the color excess $E(B-V)$ and the column density of atomic hydrogen $N$(H) where those quantities cannot be directly determined (e.g., Herbig 1993; Friedman et al. 2011; Phillips et al. 2013).

The discovery of SN 2014J in M82 (Fossey et al. 2014) provided an opportunity to obtain moderately high resolution, high signal-to-noise (S/N) ratio spectra of a bright, nearby supernova at multiple epochs (both before and after maximum apparent brightness) -- enabling searches for weak interstellar features and for temporal variations in the absorption that might reveal changes in circumstellar material (e.g., Patat et al. 2007; Simon et al. 2009; Sternberg et al. 2011, 2013).
As summarized by Goobar et al. (2014), analyses of the early photometry and optical spectra suggest that SN 2014J is a Type Ia, significantly reddened by intervening interstellar material located primarily within M82; estimates of $E(B-V)$ range from about 0.8 to 1.3 mag (Polshaw et al. 2014; Amanullah et al. 2014; Foley et al. 2014; Marion et al. 2014).
Absorption from a number of known interstellar species and from many DIBs in M82 thus might be detectable in high-S/N ratio spectra of the supernova.
Unfortunately, Type Ia SNe typically have very little UV flux below about 2600 \AA\ (e.g., Foley \& Kirshner 2011; Foley et al. 2014), making UV spectra impractical to obtain (particularly for appreciably reddened SNe) -- so that investigations of the intervening ISM will have to depend largely on interpreting the species observable in the optical and infrared.

In this paper, we discuss the interstellar absorption features seen in multi-epoch, moderately high resolution, high-S/N ratio optical spectra obtained with the ARC echelle spectrograph (ARCES; Wang et al. 2003) at Apache Point Observatory (APO).
Section 2 describes the observations and the processing of the raw spectral images.
Section 3 presents the observed interstellar features, with some comparison to those seen toward several other SNe.
Section 4 discusses some of the atomic and molecular species and several of the DIBs -- in the context of trends observed in the ISM of our Galaxy, the LMC, and the SMC -- yielding some insights into the behavior of the DIBs and the properties of the M82 ISM toward SN 2014J.
Section 5 summarizes our results and conclusions.
Two companion papers focus on the abundances and kinematics of the observed atomic and molecular species (Ritchey et al. 2014b) and on a more complete census of the DIBs detected toward SN 2014J (D. York et al., in preparation).

\section{OBSERVATIONS AND DATA ANALYSIS}
\label{sec-obs}

Forty-seven 20-minute exposures of SN 2014J were obtained with ARCES on six nights (2014 January 27, 30; February 08, 11, 22; March 04) -- i.e., both before and after the observed peak in V-band brightness of the SN around 2014 February 03.
With the currently installed echelle grating, the spectra cover nearly the full optical range from about 3800 to 11000 \AA\ at a resolution of about 31,500 (slightly lower than with the previous grating).
Independent reductions of the individual raw spectral images, employing standard procedures for ARCES spectra (e.g., Thorburn et al. 2003), were performed by AR and JD. 
Minor differences in the extracted spectra for any given individual exposure may be attributed to slight differences employed in removing cosmic ray events.
The 5--13 individual spectra obtained on each night were corrected for telluric absorption features (via observations of bright, lightly-reddened standard stars), shifted to a common velocity scale (local standard of rest), and co-added to produce nightly sum spectra.
Because no significant changes were noted in the absorption-line profiles (Ritchey et al. 2014b), a total sum of all 47 spectra was also constructed (though only the 29 spectra from the first four nights were used below about 4500 \AA, due to the significantly lower flux in the blue on the last two nights).

%%%%%%%%%%
\begin{figure}
\epsscale{0.9}
%\plotone{spec.eps}
\plotone{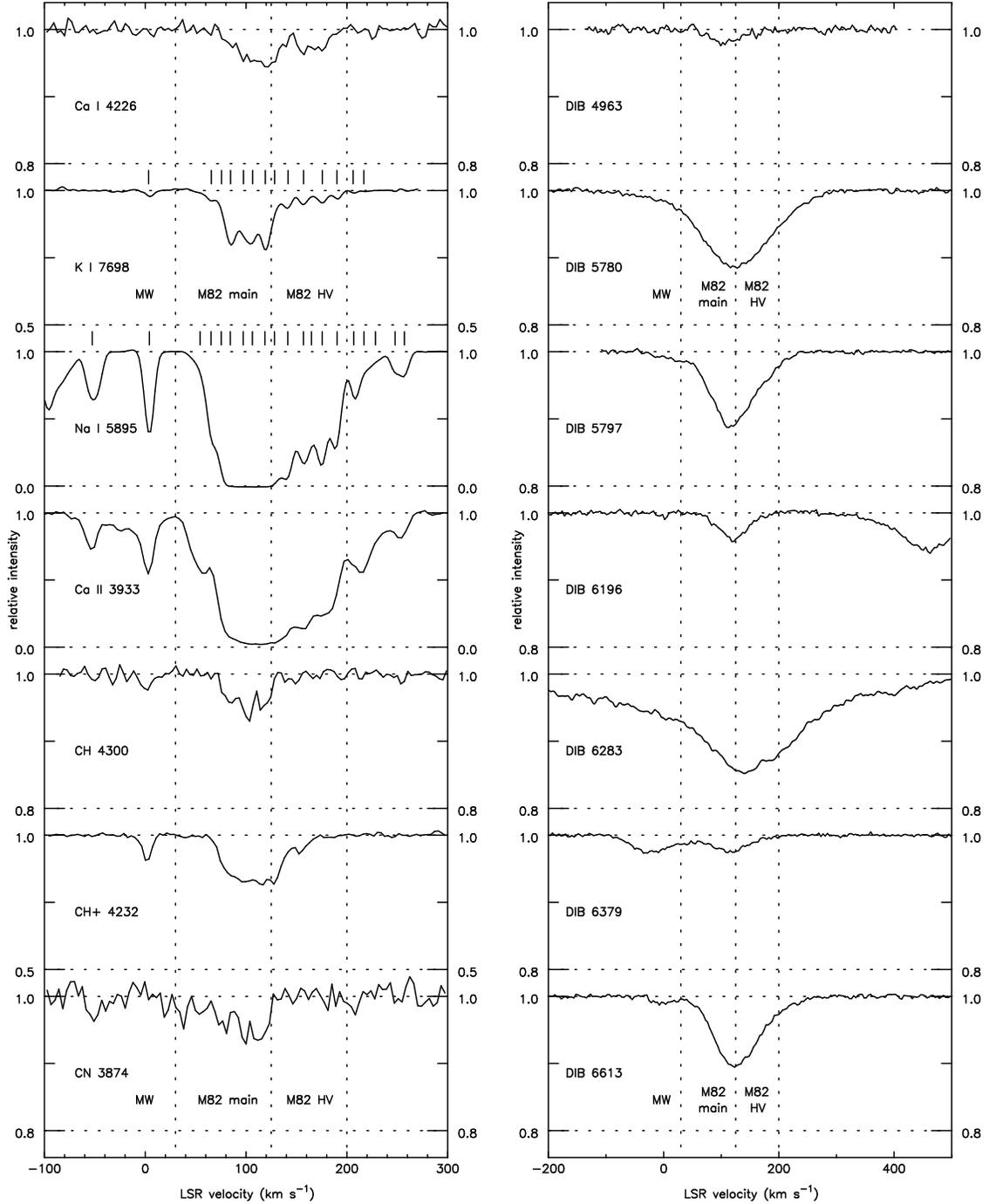}
\caption{ARCES spectra of interstellar atomic and molecular absorption lines (left) and selected diffuse interstellar bands (right) toward SN 2014J.
Individual components found in fits to the \ion{K}{1} and \ion{Na}{1} profiles are noted by tick marks.
Dotted vertical lines separate three groups of \ion{K}{1} components:  the Milky Way (MW) components (at $v_{\rm LSR}$ $\la$ 30 km~s$^{-1}$), the stronger ''main'' M82 components between 85 and 120 km~s$^{-1}$ and the weaker, higher velocity (HV) M82 components between 125 and 190 km~s$^{-1}$; note the difference in velocity scales between the left and right panels.
Still weaker M82 components are seen in \ion{Na}{1} and \ion{Ca}{2}.
The absorption between about 10 and 60 km~s$^{-1}$ for CN is for the CN R1 line.}
\label{fig:prof}
\end{figure}
%%%%%%%%%%

Spectral segments around the various interstellar absorption features of interest were normalized via low-order polynomial fits to adjacent continuum regions.
The nightly sum spectra have S/N ratios of order 300--400 (per resolution element) near 6000 \AA, and of order 100 below 4000 \AA, as determined from the scatter in the continuum fits.
Equivalent widths and apparent optical depth (AOD) estimates for the total column densities of known atomic and molecular species and equivalent widths for the DIBs were then measured from the normalized spectra by integration over the line profiles.
Where possible, separate values were measured for the absorption from our Galaxy and from M82, taking $v_{\rm LSR}$ = 30 km~s$^{-1}$ as the dividing line.
For the DIBs, the ranges over which the equivalent widths were measured are consistent with those adopted by Friedman et al. (2011; see their figure 1), with allowance for broadening of the features due to the complex component structure observed for the material in M82.
The uncertainties in the measured equivalent widths include contributions from photon noise, continuum placement, and the range in values found for different combinations of the two independently reduced sets of spectra.
Independent multi-component fits to the profiles of the atomic and molecular lines were performed, using the codes {\bf ismod} (AR; see Sheffer et al. 2008) and {\bf fits6p} (DW; see Welty et al. 2003), to obtain estimates for column densities, line widths, and velocities for the ''individual'' components discernible in the spectra (see Ritchey et al. 2014b for details).
The components found in fits to the \ion{K}{1} and \ion{Na}{1} lines, for example, are noted by tick marks above the profiles in Figure~\ref{fig:prof}. 
The total column densities obtained in the profile fits are consistent with the AOD estimates (which represent lower limits where some portion of the profile is saturated).

%%%%%  Table 1  %%%%%
\begin{deluxetable}{rcccccc}
\tablecolumns{7}
\tabletypesize{\scriptsize}
\tablecaption{M82 DIBs \label{tab:dibs}}
\tablewidth{0pt}

\tablehead{
\multicolumn{1}{c}{DIB}&
\multicolumn{1}{c}{FWHM\tablenotemark{a}}&
\multicolumn{1}{c}{$W$(total)}&
\multicolumn{1}{c}{$W$(MW)\tablenotemark{b}}&
\multicolumn{1}{c}{$W$(MW)\tablenotemark{c}}&
\multicolumn{1}{c}{$W$(MW)\tablenotemark{d}}&
\multicolumn{1}{c}{$W$(M82)\tablenotemark{e}}\\
\multicolumn{1}{c}{}&
\multicolumn{1}{c}{(\AA)}&
\multicolumn{1}{c}{(meas)}&
\multicolumn{1}{c}{(pred H/K I)}&
\multicolumn{1}{c}{(meas)}&
\multicolumn{1}{c}{(adopt)}&
\multicolumn{1}{c}{(adopt)}\\
\multicolumn{1}{c}{(1)}&
\multicolumn{1}{c}{(2)}&
\multicolumn{1}{c}{(3)}&
\multicolumn{1}{c}{(4)}&
\multicolumn{1}{c}{(5)}&
\multicolumn{1}{c}{(6)}&
\multicolumn{1}{c}{(7)}}

\startdata
%% DIB    FWHM     meas        MW       MWmeas        adopt        M82  
4963.9  & 0.62 &  23$\pm$3 &   2/1   &  \nodata  &   (1)      &  23$\pm$3  \\
5487.7  & 5.20 &  96$\pm$13&  14/14  &  \nodata  &   7$\pm$4  &  89$\pm$13 \\
5705.1  & 2.58 &  61$\pm$7 &  12/13  &  \nodata  &   6$\pm$3  &  55$\pm$8  \\
5780.5  & 2.11 & 343$\pm$13&  46/50  & 18$\pm$3  &  24$\pm$12 & 319$\pm$18 \\
5797.1  & 0.77 & 218$\pm$4 &  14/16  &  7$\pm$2  &   8$\pm$4  & 210$\pm$6  \\
6196.0  & 0.42 &  46$\pm$5 &   5/5   &  2$\pm$1  &   (2)      &  46$\pm$5  \\
6203.6  & 4.87 & 162$\pm$13&  24/24  &  \nodata  &  12$\pm$6  & 150$\pm$14 \\
6283.8  & 4.77 & 956$\pm$16& 173/188 &  \nodata  &  90$\pm$45 & 866$\pm$48 \\
6379.3  & 0.58 &  41$\pm$4 &   5/5   &  \nodata  &   (3)      &  41$\pm$4  \\
6613.6  & 0.93 & 226$\pm$5 &  10/10  &  $<$9     &   (5)      & 226$\pm$5  \\
\enddata
\tablenotetext{a}{FWHM measured toward HD~204827 (Hobbs et al. 2009).}
\tablenotetext{b}{Predicted $W$(DIB) for Milky Way, assuming log[$N$(H)] = 20.56 and Galactic relationships from Friedman et al. (2011) and assuming log[$N$(\ion{K}{1})] = 10.52 and Galactic relationships given in Appendix Table~\ref{tab:corr}.}
\tablenotetext{c}{Measured $W$(DIB) for Milky Way, for narrower DIBs.}
\tablenotetext{d}{Adopted $W$(DIB) for Milky Way:  0.5 $\times$ the predicted values, with $\pm$50\% assumed uncertainty.}
\tablenotetext{e}{$W$(DIB) for M82:  measured total minus adopted Galactic (for broader DIBs) or measured M82 value (for narrower DIBs).}
\end{deluxetable}
%%%%%%%%%%

\section{RESULTS}
\label{sec-res}

Absorption from \ion{Na}{1}, \ion{K}{1}, \ion{Ca}{1}, \ion{Ca}{2}, CH, CH$^+$, CN, and many of the DIBs is clearly visible in the summed, normalized spectra of SN 2014J (Fig.~\ref{fig:prof}).
The atomic and molecular lines exhibit multiple components, at LSR velocities ranging from about $-$53 to +257 km~s$^{-1}$ (Cox et al. 2014; Goobar et al. 2014; Ritchey et al. 2014a, 2014b; Foley et al. 2014).
Examination of the available 21 cm emission profiles in the region around M82 (e.g., Kalberla et al. 2005) suggests that the components at $v_{\rm LSR}$ $\la$ 30 km~s$^{-1}$ are due to gas in the Galactic disk and halo, with total $N$(H) $\sim$ 3.65 $\times$ 10$^{20}$ cm$^{-2}$, while components at higher velocities are due to gas associated with M82.
For \ion{K}{1} and CH, the weak absorption near $v_{\rm LSR}$ = 0 km~s$^{-1}$ is due to the main Galactic disk component (seen more strongly in \ion{Na}{1} and \ion{Ca}{2}); several relatively strong components from about 85 to 120 km~s$^{-1}$ (saturated in \ion{Na}{1}) and weaker components from about 65 to 80 and 125 to 190 km~s$^{-1}$ arise in M82; additional components, at both lower and higher velocities, are visible in the stronger lines of \ion{Na}{1} and \ion{Ca}{2} (Ritchey et al. 2014b).
That complex structure is not directly discernible in the intrinsically broader DIBs, where the contributions from the individual components seen in \ion{K}{1} overlap and blend together to produce relatively smooth, broad absorption features (though see Sec.~\ref{sec-shift} below).

While many DIBs are detected toward SN 2014J (D. York et al., in preparation), we focus in this paper on a subset of ten (Table~\ref{tab:dibs}):  the eight relatively strong DIBs examined by Friedman et al. (2011), the $\lambda$6379.3 DIB, and the $\lambda$4963.9 ''C$_2$-DIB'' (Thorburn et al. 2003), for which significant sets of uniformly measured Galactic data are available for comparison.
The relatively small velocity difference between the Galactic and M82 absorption toward SN 2014J means that contributions from the two galaxies will be blended for the intrinsically broader DIBs.
The Galactic contributions to the measured total equivalent widths of those broader DIBs may be estimated via the observed Galactic column densities of \ion{K}{1} and atomic hydrogen and the mean relationships between those neutral atomic species and the DIBs found in the Galactic ISM (column 4 of the table; see Friedman et al. 2011 and the Appendix to this paper).
Comparisons with the Galactic equivalent widths measured for several of the narrower DIBs and with the residuals from crude fits to the profiles of several of the broader DIBs (which included only M82 components; see Sec.~\ref{sec-shift} below) in column 5, however, suggest that those estimated Galactic contributions are a factor of $\sim$2 too large.
Our adopted estimates for the Galactic contributions to the DIBs, with assumed $\pm$50\% uncertainties, are given in column 6 of the table; those uncertainties are included in the uncertainties listed for the derived host galaxy equivalent widths in column 7.
As we find no significant temporal variations in the strengths of the interstellar features toward SN 2014J (Ritchey et al. 2014b), differences between the DIB equivalent widths listed in Table~\ref{tab:dibs} and values reported in several other early studies of the SN presumably reflect differences in continuum placement, integration limits, and/or the inclusion/exclusion of Galactic contributions.

Table~\ref{tab:ewn} compares the host galaxy column densities for the atomic and molecular species and the host galaxy equivalent widths for the ten selected DIBs found toward SN 2014J with those found toward four other extragalactic SNe and in three sight lines in our Galaxy.
Of the four other extragalactic sight lines listed in columns 3--6 of the table, those toward SN 1986G (NGC 5128; D'Odorico et al. 1989), SN 2006X (M100; Lauroesch et al. 2006; Patat et al. 2007; Cox \& Patat 2008), and SN 2008fp (ESO428-G14; Cox \& Patat 2014) have among the highest $A_{\rm V}$ and $E(B-V)$ in the Phillips et al. (2013) sample of 32 Type Ia SNe, and also exhibit absorption from CH, CH$^+$, and/or CN.
The values for the host galaxy $A_{\rm V}$ and $R_{\rm V}$ listed by Phillips et al. (2013) imply corresponding host galaxy $E(B-V)$ $\sim$ 0.8, 1.4, and 0.6, respectively, toward those three SNe.
No molecular absorption was detected toward SN 2001el (NGC 1448; Sollerman et al. 2005; this paper), but relatively strong absorption from \ion{Na}{1} and \ion{Ca}{2} and measurable absorption from a number of DIBs have been noted; the estimated $E(B-V)$ is 0.28.
[While in principle $A_{\rm V}$ should be a better indicator of the total amount of dust, we also consider $E(B-V)$, for which well determined values are available for more sight lines covering a wider dynamic range -- both for our Galaxy and for the Magellanic Clouds.]

The column densities and DIB equivalent widths found toward the three stars in our Galaxy are given in columns 7--9 of Table~\ref{tab:ewn}.
The sight line toward HD~62542 is dominated by a single cloud with high molecular fraction (Snow et al. 2002); \'{A}d\'{a}mkovics et al. 2005).
The sight lines toward the two DIB atlas stars HD~183143 and HD~204827 exhibit distinctly different relative molecular abundances and DIB strengths (Thorburn et al. 2003; Hobbs et al. 2008, 2009), and may have comparable $E(B-V)$ to SN 2014J.
The high abundances of CN, C$_2$, and C$_3$ toward HD~62542 and HD~204827 suggest that the densities and molecular fractions are high in the main clouds in those sight lines; the lower $N$(CN)/$N$(CH) and higher $N$(CH$^+$)/$N$(CH) ratios toward HD~183143 suggest that the clouds there are more diffuse.
Toward HD~183143, nine of the ten DIBs listed in the table lie on or above the general Galactic trends versus $E(B-V)$ and $N$(\ion{K}{1}); toward HD~204827, nine of the ten DIBs lie on or below those general trends, with only the $\lambda$4963.9 C$_2$-DIB stronger than average. 
The much lower $W$(5797.1)/$W$(5780.5) ratio toward HD~183143 suggests that the clouds in that sight line are exposed to stronger radiation fields and/or that the DIBs are less shielded (e.g., Vos et al. 2011).
The last column of the table gives average Galactic values of the various quantities, obtained from fits to the trends of those quantities versus $E(B-V)$, for sight lines with $E(B-V)$ $\sim$ 1.0.

%%%%% Table 2 %%%%%
\begin{deluxetable}{lccccccccc}
\rotate
\tablecolumns{10}
\tabletypesize{\scriptsize}
\tablecaption{Column Densities and DIB Equivalent Widths toward Five SNe and Three Galactic Stars \label{tab:ewn}}
\tablewidth{0pt}

\tablehead{
\multicolumn{1}{c}{Quantity}&
\multicolumn{1}{c}{2014J\tablenotemark{a}}&
\multicolumn{1}{c}{1986G}&
\multicolumn{1}{c}{2006X}&
\multicolumn{1}{c}{2008fp}&
\multicolumn{1}{c}{2001el}&
\multicolumn{1}{c}{HD 62542}&
\multicolumn{1}{c}{HD 183143}&
\multicolumn{1}{c}{HD 204827}&
\multicolumn{1}{c}{MW avg}\\
\multicolumn{1}{c}{ }&
\multicolumn{1}{c}{M82}&
\multicolumn{1}{c}{NGC5128}&
\multicolumn{1}{c}{M100}&
\multicolumn{1}{c}{ESO428-G14}&
\multicolumn{1}{c}{NGC1448}&
\multicolumn{1}{c}{MW}&
\multicolumn{1}{c}{MW}&
\multicolumn{1}{c}{MW}\\
\multicolumn{1}{c}{(1)}&
\multicolumn{1}{c}{(2)}&
\multicolumn{1}{c}{(3)}&
\multicolumn{1}{c}{(4)}&
\multicolumn{1}{c}{(5)}&
\multicolumn{1}{c}{(6)}&
\multicolumn{1}{c}{(7)}&
\multicolumn{1}{c}{(8)}&
\multicolumn{1}{c}{(9)}&
\multicolumn{1}{c}{(10)}}

\startdata
%%%                  2014J           1986G           2006X          2008fp          2001el            62542         
%%%   183143          204827             MW avg
$E(B-V)$\tablenotemark{b} 
                & 1.2$\pm$0.1   & 0.79$\pm$0.08 & 1.44$\pm$0.13  & 0.59$\pm$0.12  & 0.28$\pm$0.06  & 0.35$\pm$0.03  
     & 1.27$\pm$0.06 & 1.11$\pm$0.06 & (1.00)   \\
R$_{\rm V}$     & 1.6$\pm$0.2   & 2.57$\pm$0.22 & 1.31$\pm$0.09  & 1.20$\pm$0.20  & 2.25$\pm$0.41  & 2.8$\pm$0.3
     & 2.98$\pm$0.30 & 2.36$\pm$0.30 & (3.1)    \\
$A_{\rm V}$     & 1.9$\pm$0.2   & 2.03$\pm$0.11 & 1.88$\pm$0.11  & 0.71$\pm$0.09  & 0.62$\pm$0.08  & 0.99$\pm$0.14
     & 3.78$\pm$0.39 & 2.62$\pm$0.34 & (3.1)    \\
\hline
log[$N$(Na I)]  & 14.25$\pm$0.04&  13.74        &  13.8--14.3    & 14.43$\pm$0.03 & 12.76$\pm$0.03 & 13.93$\pm$0.05 
     & 14.28$\pm$0.03&  14.7$\pm$0.1 & 14.59  \\
log[$N$(K I)]   & 12.28$\pm$0.02& \nodata       & 12.00$\pm$0.04 & 12.03$\pm$0.03 & 11.29$\pm$0.05 & 11.90$\pm$0.18 
     & 12.21$\pm$0.02& 12.81$\pm$0.03& 12.57  \\
log[$N$(Ca I)]  & 11.27$\pm$0.04& \nodata       &  10.8$\pm$0.2  & 11.35$\pm$0.02 &[10.21$\pm$0.11]& $<$9.37        
     & 10.56$\pm$0.02& 10.58$\pm$0.02& 10.68  \\
log[$N$(Ca II)] & 13.83$\pm$0.05&  13.32        &[13.24$\pm$0.08]& 13.39$\pm$0.02 & 12.79$\pm$0.08 & 11.96$\pm$0.04 
     & \nodata       & $>$12.59      & \nodata\\
log[$N$(CN)]    & 13.00$\pm$0.08& \nodata       & 14.15$\pm$0.15 & [13.6$\pm$0.1] &[$<$12.05]      & 13.55$\pm$0.04 
     & 12.55$\pm$0.05& 13.74$\pm$0.04& \nodata\\
log[$N$(CH)]    & 13.71$\pm$0.05& [13.5$\pm$0.1]& 13.78$\pm$0.03 & 13.45$\pm$0.36 &[$<$12.68]      & 13.54$\pm$0.02 
     & 13.67$\pm$0.03& 13.90$\pm$0.04& 13.77  \\
log[$N$(CH$^+$)]& 14.33$\pm$0.02& [13.6$\pm$0.1]& 13.65$\pm$0.10 & 13.23$\pm$0.22 &[$<$12.66]      & $<$11.84       
     & 13.83$\pm$0.02& 13.57$\pm$0.04& 13.76  \\
\hline
$W$(4963.9)     &   23$\pm$3    & \nodata       &  [8$\pm$5]     & \nodata        &  [7$\pm$4]     &   6$\pm$1      
     &  26$\pm$1     &  53$\pm$1     &   29   \\
$W$(5487.7)     &   89$\pm$13   & \nodata       & \nodata        & \nodata        & \nodata        & \nodata        
     & 225$\pm$14    &  68$\pm$4     &  123   \\
$W$(5705.1)     &   55$\pm$8    &   79$\pm$5    & \nodata        & \nodata        &  37$\pm$5      & \nodata        
     & 172$\pm$7     &  58$\pm$3     &  114   \\
$W$(5780.5)     &  319$\pm$18   &  335$\pm$5    & $<$72          &  81$\pm$5      & 189$\pm$3      &  31$\pm$6      
     & 761$\pm$6     & 257$\pm$4     &  490   \\
$W$(5797.1)     &  210$\pm$6    &  151$\pm$5    & \nodata        &  60$\pm$10     &  26$\pm$2      &  12$\pm$2      
     & 257$\pm$8     & 199$\pm$3     &  195   \\
$W$(6196.0)     &   46$\pm$5    &   30$\pm$15   &  14$\pm$4      &  29$\pm$6      &  15$\pm$2      &   3$\pm$1      
     &  89$\pm$2     &  42$\pm$1     &   54   \\
$W$(6203.6)     &  150$\pm$14   &  191$\pm$5    & [47$\pm$15]    & [48$\pm$9]     & 102$\pm$5      &  15$\pm$4      
     & 340$\pm$11    & 116$\pm$4     &  197   \\
$W$(6283.8)     &  866$\pm$48   & \nodata       & 177$\pm$25     & 180$\pm$30     & 500$\pm$80     &  50$\pm$20     
     &1910$\pm$30    & 518$\pm$60    & 1217   \\
$W$(6379.3)     &   41$\pm$4    &  (75$\pm$8)   & $<$8           & [17$\pm$3]     &  12$\pm$3      & \nodata        
     & 105$\pm$1     &  95$\pm$1     &   87   \\
$W$(6613.6)     &  226$\pm$5    & \nodata       & $<$18          & \nodata        &  52$\pm$3      &   8$\pm$1      
     & 332$\pm$4     & 171$\pm$3     &  241   \\
\enddata
\tablerefs{2014J (Ritchey et al. 2014b; this study); 1986G (D'Odorico et al. 1989); 2006X (Patat et al. 2007; Cox \& Patat 2008; Phillips et al. 2013); 2008fp (Cox \& Patat 2014); 2001el (Sollerman et al. 2005); HD~62542 (\'{A}d\'{a}mkovics et al. 2005; Welty et al., in preparation); HD~183143 (Friedman et al. 2011; Hobbs et al. 2009); HD~204827 (Friedman et al. 2011; Hobbs et al. 2008); values in square braces were derived for this paper from archival spectra and/or quoted equivalent widths.}
\tablenotetext{a}{Estimated Galactic contributions have been removed from measured total DIB equivalent widths:  5487.7 (7$\pm$4 m\AA), 5705.1 (6$\pm$3 m\AA), 5780.5 (24$\pm$12 m\AA), 5797.1 (8$\pm$4 m\AA), 6203.6 (12$\pm$6 m\AA), 6283.8 (90$\pm$45 m\AA); see Table~\ref{tab:dibs}.}
\tablenotetext{b}{$E(B-V)$ and $A_{\rm V}$ for SN 2014J are from Goobar et al. (2014), Amanullah et al. (2014), and Foley et al. (2014); $E(B-V)$ for other SNe are derived from the $A_{\rm V}$ and $R_{\rm V}$ tabulated by Phillips et al. (2013).}
\end{deluxetable}
%%%%%%%%%%

\clearpage

%%%%% Table 3 %%%%%
\begin{deluxetable}{lccccccccc}
\rotate
\tablecolumns{10}
\tabletypesize{\scriptsize}
\tablecaption{Ratios toward Five SNe and Three Galactic Stars \label{tab:ratios}}
\tablewidth{0pt}

\tablehead{
\multicolumn{1}{c}{Quantity}&
\multicolumn{1}{c}{2014J}&
\multicolumn{1}{c}{1986G}&
\multicolumn{1}{c}{2006X}&
\multicolumn{1}{c}{2008fp}&
\multicolumn{1}{c}{2001el}&
\multicolumn{1}{c}{HD 62542}&
\multicolumn{1}{c}{HD 183143}&
\multicolumn{1}{c}{HD 204827}&
\multicolumn{1}{c}{MW avg}\\
\multicolumn{1}{c}{ }&
\multicolumn{1}{c}{M82}&
\multicolumn{1}{c}{NGC5128}&
\multicolumn{1}{c}{M100}&
\multicolumn{1}{c}{ESO428-G14}&
\multicolumn{1}{c}{NGC1448}&
\multicolumn{1}{c}{MW}&
\multicolumn{1}{c}{MW}&
\multicolumn{1}{c}{MW}\\
\multicolumn{1}{c}{(1)}&
\multicolumn{1}{c}{(2)}&
\multicolumn{1}{c}{(3)}&
\multicolumn{1}{c}{(4)}&
\multicolumn{1}{c}{(5)}&
\multicolumn{1}{c}{(6)}&
\multicolumn{1}{c}{(7)}&
\multicolumn{1}{c}{(8)}&
\multicolumn{1}{c}{(9)}&
\multicolumn{1}{c}{(10)}}

\startdata
%%%                  2014J           1986G           2006X          2008fp          2001el            62542         
%%%   183143          204827             MW avg
$E(B-V)$        & 1.2$\pm$0.1   & 0.79$\pm$0.08 & 1.44$\pm$0.13  & 0.59$\pm$0.12  & 0.28$\pm$0.06  & 0.35$\pm$0.03  
     & 1.27$\pm$0.06 & 1.11$\pm$0.06 & (1.00)   \\
R$_{\rm V}$     & 1.6$\pm$0.2   & 2.57$\pm$0.22 & 1.31$\pm$0.09  & 1.20$\pm$0.20  & 2.25$\pm$0.41  & 2.8$\pm$0.3
     & 2.98$\pm$0.30 & 2.36$\pm$0.30 & (3.1)    \\
$A_{\rm V}$     & 1.9$\pm$0.2   & 2.03$\pm$0.11 & 1.88$\pm$0.11  & 0.71$\pm$0.09  & 0.62$\pm$0.08  & 0.99$\pm$0.14
     & 3.78$\pm$0.39 & 2.62$\pm$0.34 & (3.1)    \\
                & diffuse       & diffuse       & dense          & dense          & diffuse        & dense
     & diffuse       & dense         &          \\
\hline
log[$N$(Ca I)/$N$(K I)]
                & $-$1.01$\pm$0.05  & \nodata       & $-$1.2$\pm$0.2    & $-$0.68$\pm$0.04   & $-$1.08$\pm$0.12   & $<-$2.53       
     & $-$1.65$\pm$0.03  & $-$2.23$\pm$0.04  & $-$2.05\\
log[$N$(CH)/$N$(K I)]
                &   +1.43$\pm$0.05  & \nodata       & +1.78$\pm$0.05     &   +1.42$\pm$0.36   & $<$+1.39       & +1.64$\pm$0.18     
     &   +1.46$\pm$0.04  &   +1.09$\pm$0.05  &   +1.26\\
log[$N$(CN)/$N$(CH)]
                & $-$0.71$\pm$0.09  & \nodata       & +0.37$\pm$0.15     &   +0.15$\pm$0.37   & \nodata        & +0.01$\pm$0.04     
     & $-$1.12$\pm$0.06  & $-$0.16$\pm$0.06  & \nodata\\
log[$N$(CH$^+$)/$N$(CH)]
                &  +0.62$\pm$0.05   &  +0.10$\pm$0.14    & $-$0.13$\pm$0.10   & $-$0.22$\pm$0.42   & \nodata        & $<-$1.70       
     &  +0.16$\pm$0.04   & $-$0.33$\pm$0.06  & \nodata\\
$W$(5797.1)/$W$(5780.5)
                &  0.66$\pm$0.04&  0.45$\pm$0.02& \nodata        &  0.74$\pm$0.13 &  0.14$\pm$0.01 &  0.39$\pm$0.10 
     &  0.34$\pm$0.01&  0.77$\pm$0.02&  0.41  \\
\hline
10$^5$$W$(4963.9)/[$N$(K I)]$^{1/2}$
                &   1.7$\pm$0.2 & \nodata       &    0.8$\pm$0.5 & \nodata        &   1.6$\pm$0.9  &   0.7$\pm$0.2  
     &   2.0$\pm$0.1 &   2.1$\pm$0.1 &   1.8  \\
10$^5$$W$(5780.5)/[$N$(K I)]$^{1/2}$
                &  23.1$\pm$1.4 & \nodata       & $<$7.2         &   7.8$\pm$0.6  &  42.8$\pm$2.6  &   3.5$\pm$1.0  
     &  59.8$\pm$1.5 &  10.1$\pm$0.4 &  33.9  \\
10$^5$$W$(5797.1)/[$N$(K I)]$^{1/2}$
                &  15.2$\pm$0.6 & \nodata       & \nodata        &   5.8$\pm$1.0  &   5.9$\pm$0.6  &   1.3$\pm$0.4  
     &  20.2$\pm$0.8 &   7.8$\pm$0.3 &  13.0  \\
10$^5$$W$(6196.0)/[$N$(K I)]$^{1/2}$
                &   3.3$\pm$0.4 & \nodata       &    1.4$\pm$0.4 &   2.8$\pm$0.6  &   3.4$\pm$0.5  &   0.3$\pm$0.1  
     &   7.0$\pm$0.2 &   1.7$\pm$0.1 &   3.7  \\
10$^5$$W$(6283.8)/[$N$(K I)]$^{1/2}$
                &  62.7$\pm$3.8 & \nodata       &   17.7$\pm$2.6 &  17.4$\pm$3.0  & 113.2$\pm$19.3 &   5.6$\pm$2.5  
     & 150.0$\pm$4.2 &  20.4$\pm$2.5 &  93.3  \\
\hline
$W$(4963.9)/$E(B-V)$ 
                &   19$\pm$3    & \nodata       &    6$\pm$4     & \nodata        &   25$\pm$15    &   17$\pm$3     
     &   20$\pm$1    &   48$\pm$3    &   25   \\
$W$(5780.5)/$E(B-V)$ 
                &  266$\pm$27   &  424$\pm$43   & $<$50          &   137$\pm$29   &  675$\pm$145   &   89$\pm$19    
     &  599$\pm$29   &  232$\pm$13   &  463   \\
$W$(5797.1)/$E(B-V)$
                &  175$\pm$15   &  191$\pm$20   & \nodata        &   102$\pm$27   &   93$\pm$21    &   34$\pm$6     
     &  202$\pm$11   &  179$\pm$10   &  182   \\
$W$(6196.0)/$E(B-V)$
                &   38$\pm$5    &   38$\pm$19   &   10$\pm$3     &    49$\pm$14   &   54$\pm$14    &    9$\pm$3     
     &   70$\pm$4    &   38$\pm$3    &   52   \\
$W$(6283.8)/$E(B-V)$
                &  722$\pm$72   & \nodata       &  123$\pm$21    &   305$\pm$80   & 1786$\pm$478   &  143$\pm$58    
     & 1504$\pm$75   &  467$\pm$60   & 1151   \\
\hline
$W$(4963.9)/$A_{\rm V}$
                &   12$\pm$2    & \nodata       &    4$\pm$3     & \nodata        &   11$\pm$7     &    6$\pm$1
     &    7$\pm$1    &   20$\pm$3    &   8    \\
$W$(5780.5)/$A_{\rm V}$
                &  168$\pm$20   &  165$\pm$9    & $<$38          &   114$\pm$16   &  305$\pm$40    &   31$\pm$8
     &  201$\pm$21   &   98$\pm$13   & 150    \\
$W$(5797.1)/$A_{\rm V}$
                &  111$\pm$12   &   74$\pm$5    & \nodata        &    85$\pm$18   &   42$\pm$6     &   12$\pm$3
     &   68$\pm$7    &   76$\pm$10   &  58    \\
$W$(6196.0)/$A_{\rm V}$
                &   24$\pm$4    &   15$\pm$7    &    7$\pm$2     &    41$\pm$10   &   24$\pm$4     &    3$\pm$1
     &   24$\pm$2    &   16$\pm$2    &  17    \\
$W$(6283.8)/$A_{\rm V}$
                &  456$\pm$54   & \nodata       &   94$\pm$14    &   254$\pm$53   &  806$\pm$166   &   51$\pm$21
     &  505$\pm$53   &  198$\pm$34   & 331    \\
%%%                 2014J           1986G           2006X           2008fp           2001el           62542          
%%%    183143          204827         MW avg
\enddata
\tablecomments{Based on column densities and equivalent widths listed in Table~\ref{tab:ewn}.}
\end{deluxetable}
%%%%%%%%%%

\section{DISCUSSION}
\label{sec-disc}

\subsection{Trends in the Milky Way and Magellanic Clouds}
\label{sec-local}

\begin{figure}
\epsscale{1.0}
%\plotone{k1dib_ebv.eps}
\plotone{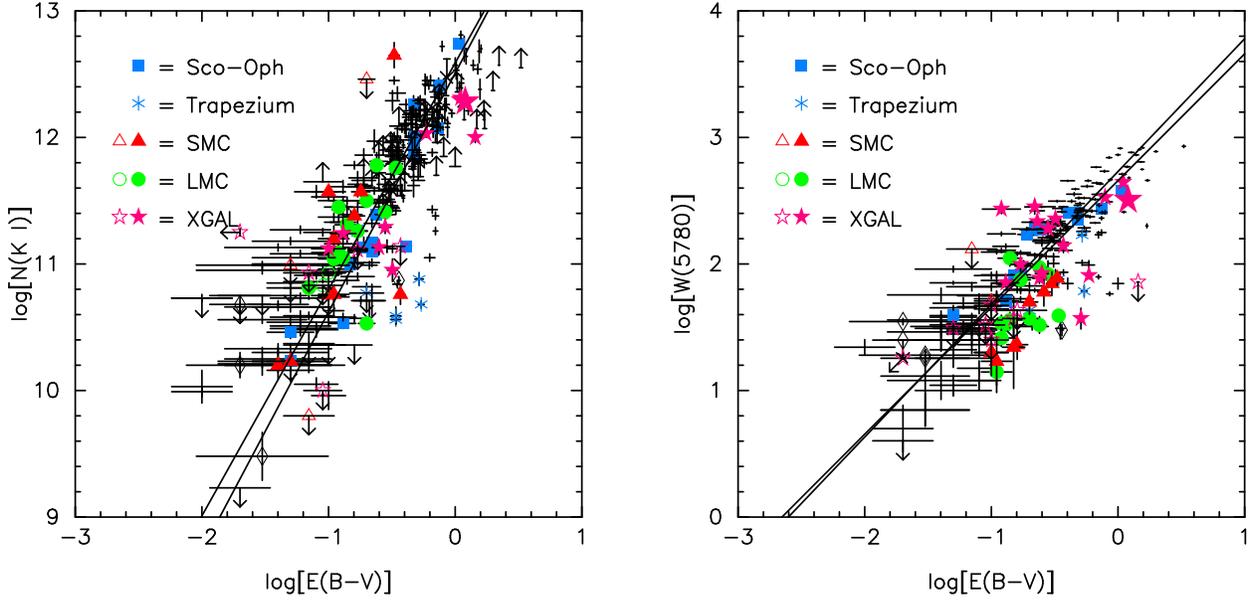}
\caption{$N$(\ion{K}{1}) and $W$(5780.5) versus $E(B-V)$, for both Galactic and extragalactic sight lines.
Solid symbols for LMC, SMC, and extragalactic (XGAL) sight lines denote detections; open symbols denote limits.
The Galactic data (plain black crosses) are mostly from Welty \& Hobbs (2001) and Friedman et al. (2011); the LMC and SMC data are mostly from Welty et al. (2006) and Welty \& Crowther (in preparation).
Extragalactic sight lines include quasar absorption-line systems, M31, M33, and external galaxies probed by SNe (mostly from Phillips et al. 2013; see text for other references); the larger XGAL star denotes the M82 gas toward SN 2014J.
The solid lines show weighted and unweighted fits to the Galactic sight lines (not including the Sco-Oph and Trapezium sight lines).
The relationship for \ion{K}{1} is nearly quadratic (slope $\sim$ 1.8--1.9); the relationship for the $\lambda$5780.5 DIB is roughly linear (slope $\sim$ 1.0).
Some sight lines with lower than typical $N$(\ion{K}{1}) have enhanced radiation fields; most of the sight lines with lower than typical $W$(5780.5) have strong CN and/or higher molecular fractions.}
\label{fig:kdvsebv}
\end{figure}

\begin{figure}
\epsscale{1.0}
%\plotone{dibs_k1.eps}
\plotone{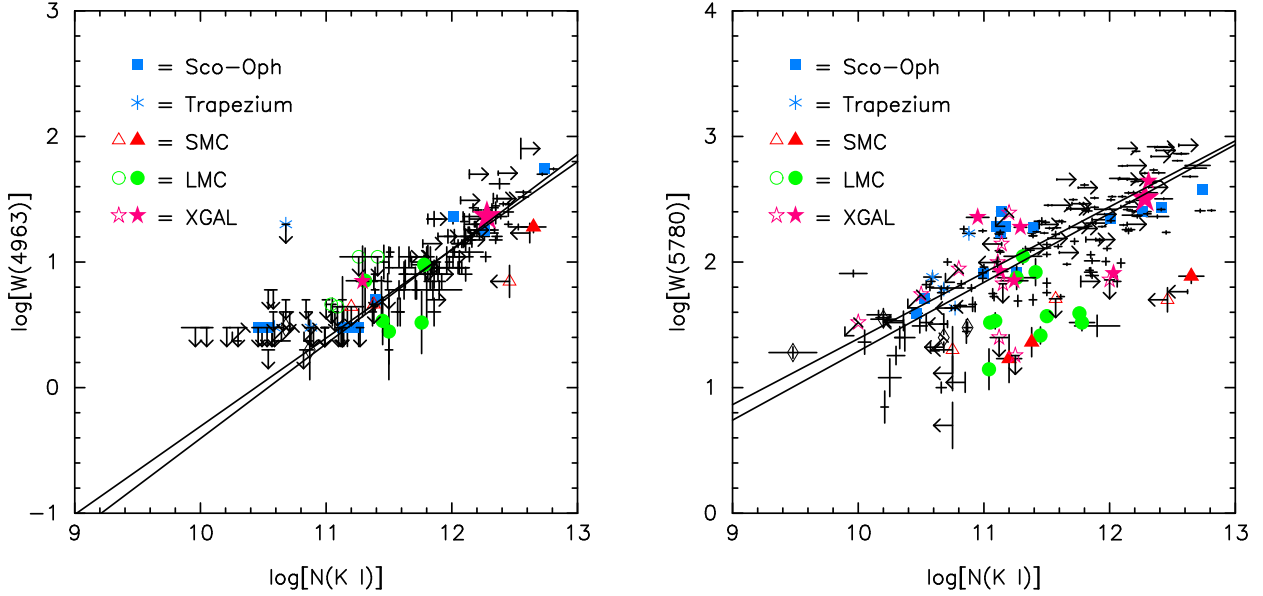}
\caption{Equivalent widths of $\lambda$4963.9 and $\lambda$5780.5 DIBs versus $N$(\ion{K}{1}), for both Galactic and extragalactic sight lines.
Solid symbols for LMC, SMC, and extragalactic (XGAL) sight lines denote detections; open symbols denote limits.
The Galactic data (plain black crosses) are mostly from Welty \& Hobbs (2001), Thorburn et al. (2003), and Friedman et al. (2011); the LMC and SMC data are mostly from Welty et al. (2006) and Welty \& Crowther (in preparation); the extragalactic data are mostly from Phillips et al. (2013).
The solid lines show weighted and unweighted fits to the Galactic sight lines (not including the Sco-Oph and Trapezium sight lines).
The relationship for the $\lambda$4963.9 DIB (a C$_2$-DIB) is steeper (slope $\sim$ 0.7), with less scatter, than the relationship for the $\lambda$5780.5 DIB (slope $\sim$ 0.5).
The larger scatter for the $\lambda$5780.5 DIB may be related to variations in metallicity and/or molecular fraction; detailed comparisons indicate that sight lines deficient in $\lambda$5780.5 vs. \ion{K}{1} tend also to be deficient vs. $E(B-V)$ (see Figs~\ref{fig:kdvsebv} and \ref{fig:resid}).
The M82 gas toward SN 2014J (larger XGAL star near log[$N$(\ion{K}{1})] = 12.3) lies slightly below the general Galactic trend for $\lambda$5780.5.}
\label{fig:dibvsk1}
\end{figure}

In the local Galactic ISM, a number of fairly well defined trends may be noted (pairwise) between the column densities of \ion{Na}{1}, \ion{K}{1}, H, and H$_2$, the equivalent widths of various DIBs, and $E(B-V)$ (Herbig 1993; Welty \& Hobbs 2001; Friedman et al. 2011). 
Unusual environmental conditions characterizing particular regions or individual sight lines may account for observed deviations from those general trends.
For example, the column densities of the trace neutral species \ion{K}{1} and \ion{Na}{1} exhibit nearly quadratic relationships with the total hydrogen column density $N$(H$_{\rm tot}$) = $N$(H) + 2$N$(H$_2$) and with $E(B-V)$ and $A_{\rm V}$ (left-hand panel of Fig.~\ref{fig:kdvsebv}) -- consistent with considerations of ionization equilibrium -- and roughly linear relationships with $N$(H$_2$) [for $N$(H$_2$) $>$ 10$^{18}$ cm$^{-2}$] (Welty \& Hobbs 2001)\footnotemark.
\footnotetext{Unless otherwise stated, column densities for Galactic and Magellanic Clouds sight lines used in the correlation plots are from online compilations maintained at http://astro.uchicago.edu/$\sim$dwelty/coldens.html and coldens\_mc.html, where references to the sources of the data may also be found.}
The DIBs, on the other hand, appear to exhibit roughly linear relationships with $N$(H), $N$(H$_{\rm tot}$), $E(B-V)$, and $A_{\rm V}$ (right-hand panel of Fig.~\ref{fig:kdvsebv}), and roughly square-root relationships with $N$(\ion{Na}{1}) and $N$(\ion{K}{1}) (i.e., slopes $\sim$0.5 in log-log plots; Fig.~\ref{fig:dibvsk1}) (Herbig 1993; Welty et al. 2006; Friedman et al. 2011; Welty 2014).
While for many DIBs there is no significant residual or secondary correlation with $N$(H$_2$) (Herbig 1993; Friedman et al. 2011), there have been indications that some DIBs can be affected (positively or negatively) by the presence of molecular gas (Kre{\l}owski et al. 1999; Weselak et al. 2004, 2008; Lan et al. 2014); in particular, the relatively weak, narrow ''C$_2$-DIBs'' generally are enhanced where C$_2$ and CN are abundant (Thorburn et al. 2003; Welty 2014).
These general Galactic relationships motivate the ratios given in Table~\ref{tab:ratios} -- $W$(DIB)/$E(B-V)$, $W$(DIB)/$A_{\rm V}$, and $W$(DIB)/[$N$(\ion{K}{1})]$^{1/2}$ -- which can facilitate the comparisons of DIB strengths in different sight lines.
[Note that the DIBs generally appear to be unsaturated -- at least for $W$(5780.5) $\la$ 800 m\AA\ (e.g., Thorburn et al. 2003) -- so that their equivalent widths are linearly related to the abundances of the carrier molecules and may be directly compared with the column densities of the atomic and molecular species.]
In this paper, we focus on comparisons between the DIBs and \ion{K}{1}, as it can be difficult to gauge potential saturation effects on the strong \ion{Na}{1} $\lambda\lambda$5890.0, 5895.9 lines in the moderate-resolution, modest S/N ratio spectra typically available for extragalactic targets -- which can make it difficult to determine accurate \ion{Na}{1} column densities.\footnotemark
\footnotetext{We note that unrecognized saturation effects in the Galactic \ion{Na}{1} data tabulated by Welsh et al. (2010) and used by Phillips et al. (2013) to define fiducial Galactic relationships appear to have biased the slopes of those fiducial relationships.
The slopes for Galactic $N$(\ion{Na}{1}) and $N$(\ion{K}{1}) versus $A_{\rm V}$ (assumed to be the same) are thus underestimated ($\sim$1.1 vs. $\sim$1.8--2.0), and the slope for $W$(5780.5) versus $N$(\ion{Na}{1}) is overestimated ($\sim$0.9 vs. $\sim$0.5) there.
Correction of the fiducial relationships for those biases may affect the characterization of sight lines toward some extragalactic SNe as having ''anomalously strong'' \ion{Na}{1}.}

%%%%%%%%%%
%%%%%%%%%%

%%%%%%%%%%
%%%%%%%%%%

\begin{figure}
%\epsscale{0.7}
%\plottwo{resid1.eps}{resid2.eps}
\plottwo{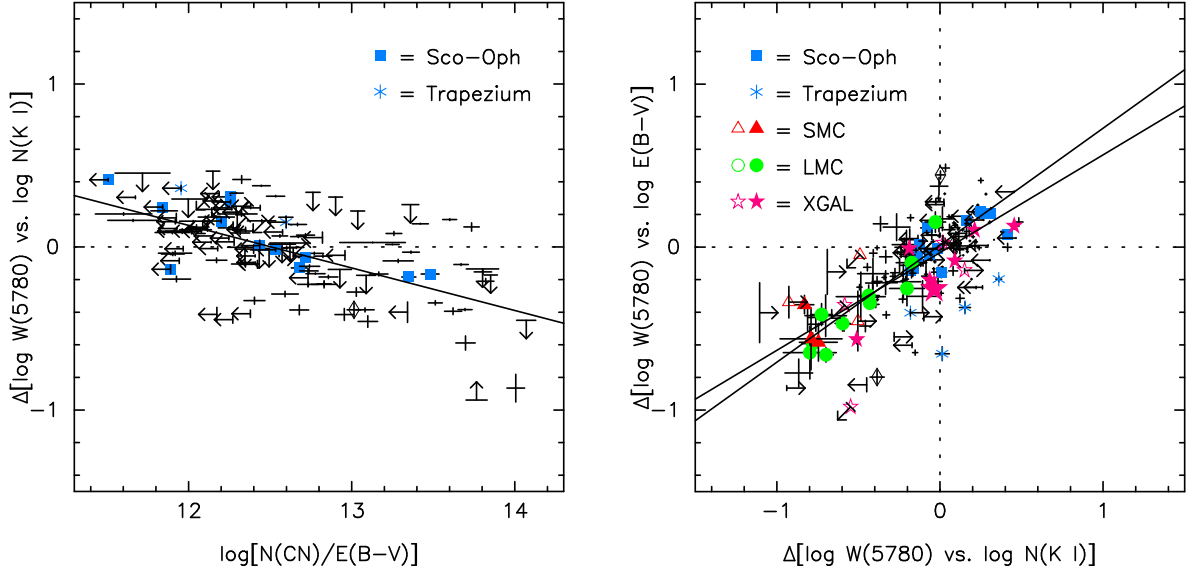}{f4b.eps}
\caption{Residuals (observed minus mean Galactic relationship) for $W$(5780.5) versus $N$(\ion{K}{1}), plotted against the ratio $N$(CN)/$E(B-V)$, for Galactic sight lines (left), and residuals for $W$(5780.5) versus $E(B-V)$ plotted against corresponding residuals for $W$(5780.5) versus $N$(\ion{K}{1}), for an expanded sample which includes LMC, SMC, and other extragalactic (XGAL) sight lines (right).
The left-hand panel indicates that the $\lambda$5780.5 DIB tends to be weaker than would be expected from $N$(\ion{K}{1}) in sight lines with strong CN absorption.
The right-hand panel indicates that the residuals versus $E(B-V)$ and $N$(\ion{K}{1}) are correlated; the best-fit lines, for weighted and unweighted fits to the Galactic data (excluding the Trapezium and Sco-Oph sight lines), have slopes $\sim$0.7.
Sight lines with a significant fraction of colder, denser molecular material thus tend to have large negative residuals with respect to the general trends for $W$(5780.5) versus both $N$(\ion{K}{1}) and $E(B-V)$.
The LMC, SMC, and other extragalactic sight lines appear to exhibit very similar behavior.
Given measurements of $N$(\ion{K}{1}) and $W$(5780.5), the relationship between the residuals can be used to refine estimates for $E(B-V)$ derived from observations of $W$(5780.5) alone -- though it does not account for the possible effects of strong local radiation fields (e.g., for the Trapezium sight lines).}
\label{fig:resid}
\end{figure}

Examination of the ''outlier'' sight lines in such correlation plots enables exploration of the effects of local physical conditions on the strengths of various interstellar species.
Lower than expected values for $N$(\ion{Na}{1}), $N$(\ion{K}{1}), and the equivalent widths of some DIBs, relative to $N$(H$_{\rm tot}$), in the Sco-Oph and (especially) the Orion Trapezium regions may be due to enhanced radiation fields (e.g., Herbig 1993; Welty et al. 2006; Vos et al. 2011).
There are a number of other sight lines in which some of the typically stronger, well-studied ''standard'' DIBs (e.g., $\lambda$5780.5 and the others examined by Friedman et al. 2011) are weaker than expected, relative to the general trends with $N$(H$_{\rm tot}$), $N$(\ion{Na}{1}), $N$(\ion{K}{1}), $E(B-V)$, and $A_{\rm V}$ [but not $N$(H)] (right-hand panels of Figs.~\ref{fig:kdvsebv} and \ref{fig:dibvsk1}). 
Many of those sight lines (e.g., X Per, HD~204827, HD~210121) are characterized by strong CN absorption and/or high molecular fractions $f$(H$_2$) = 2$N$(H$_2$)/$N$(H$_{\rm tot}$).
The left-hand panel of Figure~\ref{fig:resid} shows the residuals of the observed $W$(5780.5), with respect to the mean Galactic trend with $N$(\ion{K}{1}), versus the ratio $N$(CN)/$E(B-V)$, which is indicative of the amount of colder, denser molecular material.
There is a weak but definite tendency (slope $\sim$ $-$0.3) for the $\lambda$5780.5 DIB to be weaker than expected in sight lines where CN absorption is strong.
A plot of the residuals of the observed $W$(5780.5), with respect to the mean Galactic trends with both $N$(\ion{K}{1}) and $E(B-V)$ (right-hand panel of Figure~\ref{fig:resid}), shows that sight lines with weak (strong) $W$($\lambda$5780.5) relative to $N$(\ion{K}{1}) tend also to have weak (strong) $W$($\lambda$5780.5) relative to $E(B-V)$.
The scatter in the residuals plots likely reflects the different processes affecting the trace neutral species (ionization, depletion) and the DIB carriers (ionization, dissociation, depletion, hydrogenation, ...) in different sight lines.
Consideration of the sight lines with the highest molecular abundances (e.g., HD~29647, Walker 67, HD~62542, NGC2024-1) suggests that the relationship between the residuals could be somewhat steeper.

Similar relationships between the residuals are seen for the other DIBs in Tables~\ref{tab:dibs} and ~\ref{tab:ewn} -- particularly for the broader DIBs (see Appendix Table~\ref{tab:corr}).
As exemplified by $\lambda$4963.9, however, the C$_2$-DIBs (Thorburn et al. 2003) have somewhat steeper slopes, relative to $N$(\ion{Na}{1}) and $N$(\ion{K}{1}), have weak positive correlations with $f$(H$_2$), and thus are not weaker than expected in sight lines with higher $f$(H$_2$) (left-hand panel of Fig.~\ref{fig:dibvsk1}; Welty 2014).
The sight line toward HD~62542 is an extreme example -- with very weak standard DIBs but fairly normal C$_2$-DIBs (Snow et al. 2002; \'{A}d\'{a}mkovics et al. 2005).
Lan et al. (2014) measured the equivalent widths of 20 DIBs in stacked Sloan Digital Sky Survey spectra of stars, galaxies, and quasars, and found that the strengths of some of those DIBs can be either enhanced or reduced (relative to the primary correlation with atomic hydrogen) in the presence of molecular gas; one of the DIBs showing the strongest apparent enhancement is the $\lambda$4726.8 C$_2$ DIB.
The standard DIBs thus appear to trace primarily atomic gas, while the generally weaker C$_2$-DIBs can also be present (and enhanced) in gas with higher molecular fractions.
The standard DIBs appear weak, relative to $N$(\ion{Na}{1}), $N$(\ion{K}{1}), $E(B-V)$, and $A_{\rm V}$, in sight lines with appreciable dense molecular material at least in part because the latter trace both the atomic and the molecular gas.

%%%%%%%%%%
%%%%%%%%%%

Some differences in the behavior of the neutral species and DIBs may be noted in the Magellanic Clouds, where the overall metallicities and average dust-to-gas ratios are lower (by factors of $\sim$2--3 in the LMC and $\sim$4--5 in the SMC) and the typical radiation fields are somewhat stronger than in the local Galactic ISM (Lequeux 1989; Welty et al. 2012).
As in our Galaxy, $N$(\ion{Na}{1}) and $N$(\ion{K}{1}) appear to follow nearly quadratic relationships with $N$(H$_{\rm tot}$), but at much lower overall column densities -- reflecting the combined effects of lower metallicities, stronger radiation fields, and (perhaps) less grain-assisted recombination (Welty \& Hobbs 2001; Welty 2014; Welty \& Crowther, in preparation).
The relationships between $N$(\ion{Na}{1}), $N$(\ion{K}{1}), and $E(B-V)$ are more similar to those seen in the Galactic ISM, however, as all three of those quantities depend (to first order) on the metallicity (left-hand panel of Fig.~\ref{fig:kdvsebv}).
On average, the $\lambda$5780.5, $\lambda$5797.1, and $\lambda$6283.8 DIBs are weaker by factors of 7--9 in the LMC and $\sim$20 in the SMC, compared to Galactic sight lines with similar $N$(H) and $N$(H$_{\rm tot}$) (Cox et al. 2006, 2007; Welty et al. 2006), but are only slightly weaker than for Galactic sight lines with similar $E(B-V)$ (right-hand panel of Fig.~\ref{fig:kdvsebv}).
While the relationship between $W$(5780.5) and $N$(H) appears to be roughly linear in the LMC (as in our Galaxy), clear trends cannot yet be identified in the small sample available for the SMC (Welty et al. 2012; Welty 2014). 
In the Magellanic Clouds, the strengths of the $\lambda$5780.5, $\lambda$5797.1, and $\lambda$6283.8 DIBs generally fall below the Galactic trends versus $N$(\ion{Na}{1}) and $N$(\ion{K}{1}) (right-hand panel of Fig.~\ref{fig:dibvsk1}), with the most deficient values for both the LMC and SMC roughly a factor of 10 lower than the mean Galactic values (but comparable to the lowest observed Galactic values).
Even the unusual SMC sight line toward Sk 143 -- where a higher than typical ratio of density to radiation field may be responsible for the observed Galactic-like abundances of the trace neutral species, CH, C$_2$, C$_3$, CN, and the C$_2$-DIBs -- exhibits weak ''standard'' DIBs (Welty et al. 2006, 2013).
In the LMC and SMC, the residuals for the observed $W$(5780.5), relative to the mean Galactic relationships versus $N$(\ion{K}{1}) and $E(B-V)$, appear to follow the same trend seen in the local Galactic ISM (right panel of Fig.~\ref{fig:resid}).
While the most deficient of the Magellanic Clouds $W$(5780.5) generally are for sight lines with stronger molecular absorption, in most cases the overall molecular fractions are not as high as those for the Galactic sight lines with weaker than expected DIBs.
Additional metallicity-related effects may weaken the DIBs in the LMC and SMC.

\subsection{The ISM in M82}
\label{sec-m82}

\begin{figure}
\epsscale{1.0}
%\plotone{ratios.eps}
\plotone{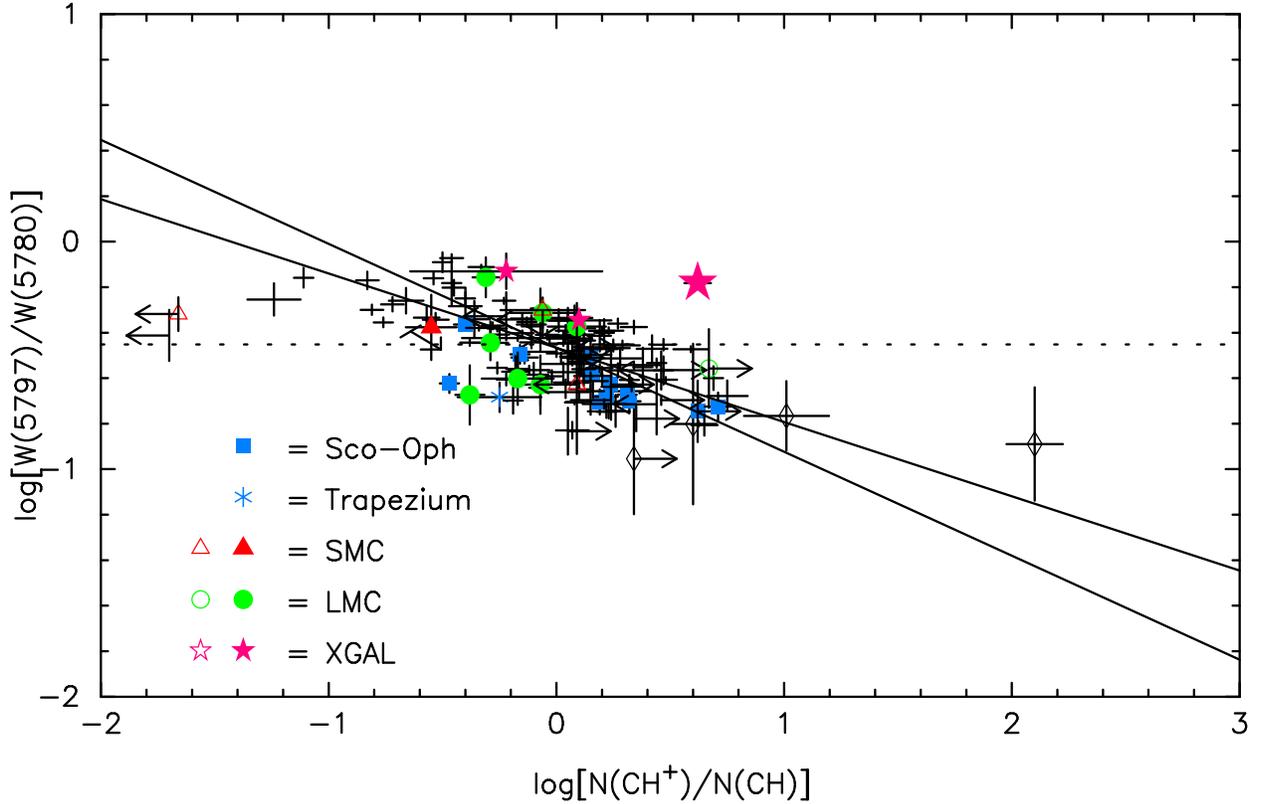}
\caption{Ratio of DIB equivalent widths $W$(5797.1)/$W$(5780.5) versus ratio of column densities $N$(CH$^+$)/$N$(CH).
The two ratios are anti-correlated in the local Galactic ISM (though points for most sight lines in the LMC and SMC, and toward SN 1986G in NGC 5128 and SN 2008fp in ESO428-G14, also are consistent); the best-fit lines have slopes $\sim$ $-$0.4.
The open black diamonds [several with large $N$(CH$^+$)/$N$(CH)] represent sight lines in the Pleiades; the sight line toward SN 2014J (larger XGAL star) exhibits a unique combination of high values for both ratios.}
\label{fig:ratios}
\end{figure}

The various absorption features observed toward SN 2014J that are due to interstellar material in M82 (Tables~\ref{tab:ewn} and \ref{tab:ratios}) exhibit both similarities and intriguing differences, when compared to the trends found in our Galaxy and in the Magellanic Clouds.
Overall, the $N$(\ion{K}{1})/$N$(\ion{Na}{1}) and $N$(CH)/$N$(\ion{K}{1}) ratios are quite consistent with the local Galactic values; the ''individual'' component values obtained from fits to the line profiles are, in most cases, also reasonably consistent (Ritchey et al. 2014b).
The $N$(\ion{Ca}{1})/$N$(\ion{K}{1}) ratio is high, especially for the higher velocity components ($v_{\rm LSR}$ $\ga$ 140 km~s$^{-1}$) -- suggestive of relatively mild depletion of calcium in M82 (Welty et al. 2003).
The general decline of the $N$(\ion{Na}{1})/$N$(\ion{Ca}{2}) ratio with increasing velocity also suggests less severe depletions in the higher velocity gas (Ritchey et al. 2014b).
While CN is detected, $N$(CN) falls well within the range observed for similar $N$(CH) in the local Galactic ISM; the $N$(CN)/$N$(CH) ratio is not as high as those seen in CN-rich Galactic sight lines (e.g., toward HD~62542 and HD~204827).
The column density of CH$^+$, however, is very high; to our knowledge, the $N$(CH$^+$)/$N$(CH) ratio in the M82 components is significantly exceeded in our Galaxy only for some sight lines in the Pleiades (e.g., Ritchey et al. 2006; Fig.~\ref{fig:ratios}).
The bulk of the M82 molecular material toward SN 2014J is thus unlikely to be in very dense clouds, and non-thermal chemical processes (needed to form the abundant CH$^+$; e.g., Zsarg\'{o} \& Federman 2003) must be quite active.

%%%%%%%%%%
%%%%%%%%%%

The DIBs in M82 seen toward SN 2014J also exhibit some similarities and some differences, relative to the general trends in the local Galactic ISM (Table~\ref{tab:ratios}).
When plotted versus $N$(\ion{K}{1}), for example, the equivalent widths of six of the ten DIBs toward SN 2014J in Table~\ref{tab:ewn} ($\lambda\lambda$ 4963.9, 5487.7, 5780.5, 6196.0, 6203.6, 6283.8) are within $\pm$20\% of the mean Galactic values at log[$N$(\ion{K}{1})] = 12.28 (Fig.~\ref{fig:dibvsk1}); the $\lambda$5705.1 and $\lambda$6379.3 DIBs are weaker by 20--45\%; and the $\lambda$5797.1 and $\lambda$6613.6 DIBs are stronger by 25--40\%.
The $W$(5797.1)/$W$(5780.5) ratio is thus fairly high ($\sim$0.7), suggestive of a relatively weak ambient radiation field and/or a somewhat shielded environment.
The combination of a high $W$(5797.1)/$W$(5780.5) ratio and a high $N$(CH$^+$)/$N$(CH) ratio is unusual, as those two ratios exhibit a fairly tight anti-correlation in the local Galactic ISM (Fig.~\ref{fig:ratios}); most of the LMC, SMC, and other extragalactic sight lines are consistent with that anti-correlation.
The slight weakness of the M82 $\lambda$5780.5 DIB, relative to the Galactic trend with $N$(\ion{K}{1}) (right-hand panel of Fig.~\ref{fig:dibvsk1}), suggests that the sight line toward SN 2014J is not dominated by cold, dense molecular gas -- consistent with indications from the molecular column densities -- and also that there are no strong metallicity-related effects on the M82 DIBs (as perhaps seen for sight lines in the SMC) -- consistent with the slightly sub-solar metallicity, [Fe/H] $\sim$ $-$0.35 dex, found for the nuclear region of M82 (Origlia et al. 2004).
The weakness of the $\lambda$6379.3 DIB toward SN 2014J was immediately suggested by comparison with the adjacent $\lambda$6376.1 DIB -- the two DIBs have similar central depths toward SN 2014J (Fig.~\ref{fig:prof}), whereas the $\lambda$6379.3 DIB is typically several times deeper than the $\lambda$6376.1 DIB in the local Galactic ISM (e.g., Galazutdinov et al. 2008b; Hobbs et al. 2008, 2009).

\subsection{Other Extragalactic Supernovae}
\label{sec-xgal}

Of the four other extragalactic sight lines in Tables~\ref{tab:ewn} and \ref{tab:ratios}, those toward SN 1986G, SN 2006X, and SN 2008fp were included because molecular species have been detected there.
Toward both SN 2006X and SN 2008fp, the $N$(CN)/$N$(CH) ratios are quite high and the $N$(CH$^+$)/$N$(CH) ratios are relatively low (Lauroesch et al. 2006; Patat et al. 2007; Cox \& Patat 2008, 2014) -- suggesting that the main host galaxy components contain relatively cold, dense gas.
For SN 2008fp, analysis of the C$_2$ rotational excitation yields a kinetic temperature of order 30 K and a density of ''collision partners'' $n_{\rm c}$ = $n$(H) + $n$(H$_2$) $\sim$ 200 cm$^{-3}$ (assuming a Galactic-like near-IR radiation field; Cox \& Patat 2014); the total hydrogen density will be somewhat larger than $n_{\rm c}$, given the apparently significant molecular fraction.
The fairly high host galaxy $N$(\ion{Ca}{1})/$N$(\ion{K}{1}) ratios toward those two SNe, normally associated with milder depletions in relatively diffuse gas, are thus somewhat unexpected.
As for SN 2014J, the host galaxy $N$(CH)/$N$(\ion{K}{1}) ratios toward the other four SNe are generally consistent with typical Galactic values.

Some differences in DIB behavior are also seen among the other extragalactic sight lines.
The sight lines toward SN 1986G (NGC 5128) and SN 2001el (NGC 1448) appear to be characterized by relatively ''normal'' DIB strengths versus both $N$(\ion{K}{1}) and $E(B-V)$, except for the weak $\lambda$5797.1 DIB toward SN 2001el (D'Odorico et al. 1989; Sollerman et al. 2005).
In the CN-rich sight lines to SN 2006X (M100) and SN 2008fp (ESO428-G14), however, the DIBs generally fall well below the Galactic trends in $W$(DIB) versus $N$(\ion{K}{1}) -- similar to the DIBs toward HD~62542.
Of the two other SNe in the Phillips et al. (2013) sample with low $W$(5780.5), relative to $N$(\ion{K}{1}) and $E(B-V)$, SN 2009ig also exhibits strong CN absorption (Cox \& Patat 2014), while no information is available for the molecular absorption toward SN 2009le.
The $\lambda$6379.3 DIB is weaker than the $\lambda$6196.0 DIB toward both SN 2006X and SN 2008fp, and is only slightly deeper than the $\lambda$6376.1 DIB toward SN 2008fp.

\subsection{Shifted DIBs in M82}
\label{sec-shift}

\begin{figure}
\epsscale{0.4}
%\plotone{dibfit.eps}
\plotone{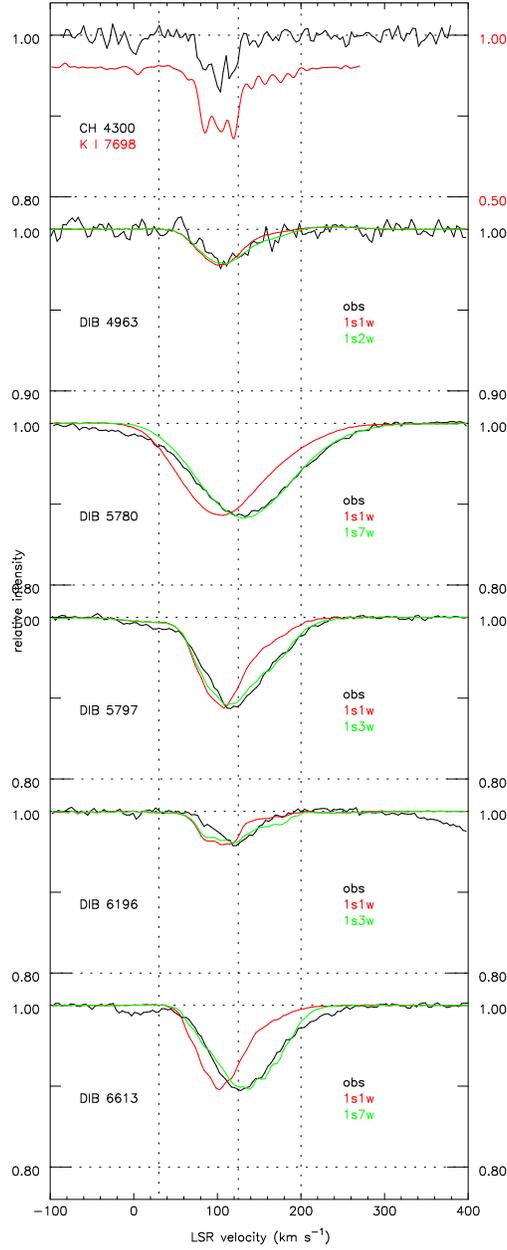}
\caption{Profiles of \ion{K}{1} $\lambda$7698, CH $\lambda$4300, and five DIBs toward SN 2014J.
The vertical dotted lines separate the three groups of components seen in \ion{K}{1}:  Milky Way; main M82; and weaker, higher velocity M82.
Note the apparent offsets in velocity between \ion{K}{1}/CH and the $\lambda\lambda$5780.5, 5797.1, 6196.0, and 6613.6 DIBs.
For the DIBs, the black line shows the observed profile; the red line shows the model profile based on weighting the \ion{K}{1} components by $N$(\ion{K}{1}) alone (using ''intrinsic'' DIB profiles from ARCES spectra of 20 Aql); and the green line shows the model profile where the weaker, higher velocity \ion{K}{1} components are weighted more heavily (by factors of 2 for $\lambda$4963.9, 3 for $\lambda$5797.1 and $\lambda$6196.0, and 7 for $\lambda$5780.5 and $\lambda$6613.6); in each case, the code ''nsmw'' notes the relative weighting factors (n, m) used for the strong and weak \ion{K}{1} component groups, respectively.
Residuals near $v_{\rm LSR}$ = 0 km~s$^{-1}$ for several of the DIBs represent contributions from the Galactic ISM.}
\label{fig:dibfit}
\end{figure}

One other notable difference in the behavior of the M82 DIBs is that the profiles of many of the DIBs appear to be shifted in velocity, relative to the envelope of the \ion{K}{1} $\lambda\lambda$7664, 7698 profiles toward SN 2014J, with different shifts for different DIBs.
For example, while the $\lambda$4963.9 DIB is fairly well aligned with the strongest \ion{K}{1} (and CH) absorption, the $\lambda\lambda$5780.5, 5797.1, and 6613.6 DIBs are shifted to the red (Fig.~\ref{fig:dibfit}).
In the Galactic ISM, the equivalent widths of the DIBs are generally correlated with $N$(\ion{K}{1}), with the narrower DIBs exhibiting the best correlations (Fig.~\ref{fig:dibvsk1}; Herbig 1993; Kre{\l}owski et al. 1998; Welty 2014).
The rest wavelengths adopted for the DIBs have therefore often been estimated via comparisons with the profiles of the \ion{K}{1} lines in relatively simple sight lines.
Combined contributions from the various components discernible in high-resolution spectra of \ion{Na}{1} and/or \ion{K}{1} for more complex sight lines can thus produce apparent shifts in the broader, unresolved DIB profiles -- e.g., for HD~183143 versus HD~204827 (Hobbs et al. 2008, 2009) or (perhaps) for some sight lines in Sco OB1 (Galazutdinov et al. 2008a) -- though not all observed DIB shifts can be explained in that way (e.g., for Herschel 36; Dahlstrom et al. 2013).

%%%%%%%%%%
%%%%%%%%%%

We have attempted to model the M82 DIB profiles toward SN 2014J by combining ''intrinsic'' DIB profiles for all the velocity components seen in \ion{K}{1} toward the SN.
The intrinsic DIB profiles, expressed as apparent optical depths as functions of velocity, were derived from ARCES spectra of 20 Aql, which has a single strong \ion{K}{1} component (at ARCES resolution) and fairly narrow observed DIBs.
Weighting the components according to their \ion{K}{1} column densities and scaling the combined profiles to the observed central depths of the DIBs yields the profiles shown in red in Figure~\ref{fig:dibfit}, which do not match the observed DIB profiles (except for $\lambda$4963.9).
Weighting the weaker, higher velocity \ion{K}{1} components (125 km~s$^{-1}$ $\la$ $v_{\rm LSR}$ $\la$ 190 km~s$^{-1}$) more heavily, however, can yield combined profiles that match the observed profiles fairly well.
For example, the green lines in Figure~\ref{fig:dibfit} show the model profiles obtained by giving higher weight to the weaker \ion{K}{1} components -- by factors of 2 for $\lambda$4963.9, 3 for $\lambda$5797.1 and $\lambda$6196.0, and 7 for $\lambda$5780.5 and $\lambda$6613.6.
Changing those weighting factors for the weaker \ion{K}{1} components by $\pm$1 generally yields somewhat poorer fits to the DIB profiles.
That simple bi-modal weighting of the two groups of \ion{K}{1} components appears to yield reasonably good matches to the observed profiles -- except for the narrow $\lambda$6196.0 DIB, for which a more complex weighting scheme may be needed.
The ''extra'' absorption near $v_{\rm LSR}$ = 0 km~s$^{-1}$ for the $\lambda\lambda$5780.5, 5797.1, and 6613.6 DIBs is likely Galactic (which was not included in the model profiles).
The factor-of-7 range in individual component $W$(5780.5)/$N$(\ion{K}{1}) implied by the differences in weighting is not unreasonable, given the observed range in total sight line values in the right-hand panel of Figure~\ref{fig:dibvsk1}.

The differences in weighting of the stronger main and weaker, higher velocity \ion{K}{1} components needed to fit the observed DIB profiles suggest that the two groups of components/clouds are characterized by different physical/environmental conditions and provide some indications as to how the different DIBs respond to those conditions.
Given the relative weights 1:3 and 1:7 noted above for the $\lambda$5797.1 and $\lambda$5780.5 DIBs and the column densities of \ion{K}{1} in the main and higher velocity component groups (14.7 and 4.2 $\times$ 10$^{11}$ cm$^{-2}$, respectively), we can estimate the equivalent widths of the two DIBs in the two component groups:  113 and 97 m\AA\ for $\lambda$5797.1 and 106 and 213 m\AA\ for $\lambda$5780.5.
For atomic hydrogen, the Galactic relationship log[$N$(H)] = 19.05 + 0.92 $\times$ log[$W$(5780.5)] (e.g., Friedman et al. 2011) and those $\lambda$5780.5 equivalent widths would imply log[$N$(H)] $\sim$ 20.92 and 21.19 cm$^{-2}$, respectively.
Similarly, the observed $N$(CH) in the two component groups (4.4 and 0.6 $\times$ 10$^{13}$ cm$^{-2}$) would imply log[$N$(H$_2$)] $\sim$ 21.08 and 20.21 cm$^{-2}$, respectively (for log[$N$(H$_2$)] = 6.82 + 1.045 $\times$ log[$N$(CH)], as seen in our Galaxy and in the LMC; e.g., Welty et al. 2006).
The main \ion{K}{1} component group would thus contain about 65\% of the total hydrogen in the sight line, with molecular fraction $f$(H$_2$) $\sim$ 0.75.
The weaker, higher velocity \ion{K}{1} component group would have about 70\% of the atomic hydrogen, with much lower $f$(H$_2$) $\sim$ 0.15.
As discussed by Ritchey et al. (2014b), the 21 cm emission in the direction of the SN (M. Yun 2014, private communication) peaks between about 100 and 160 km~s$^{-1}$; slightly more than half the M82 emission is over the velocity range corresponding to the weaker \ion{K}{1} components.

In principle, several additional factors could affect these estimates for the atomic and molecular content in the two M82 component groups.
First, the $N$(H) predicted from $W$(5780.5) could be slightly underestimated if the metallicity of M82 is slightly sub-solar (Origlia et al. 2004).
Even factor-of-2 increases in $N$(H), however, would not alter the characterization of the two component groups as significantly molecular and primarily atomic, respectively.
Second, the high CH$^+$ abundance toward SN 2014J might signal that the normal relationship between CH and H$_2$ may not apply -- as CH can be produced non-thermally along with CH$^+$ (e.g., Zsarg\'{o} \& Federman 2003) and as there are Galactic sight lines with enhanced CH$^+$ where the $N$(H$_2$)/$N$(CH) ratios can be more than an order of magnitude lower than typical Galactic values [e.g., 23 Ori (Welty et al. 1999); Herschel 36 (Dahlstrom et al. 2013)].
While the true column densities of H$_2$ (and the corresponding molecular fractions) could thus be lower, we note that the $N$(CN)/$N$(CH) ratio toward the SN is much higher than toward 23 Ori or Herschel 36 -- suggesting that at least a significant fraction of the CH seen toward the SN is due to equilibrium chemistry.

For the DIBs, the heavier weighting of the weaker \ion{K}{1} components for the $\lambda$5780.5 DIB, relative to the $\lambda$5797.1 DIB, implies that $W$(5797.1)/$W$(5780.5) $\sim$ 0.45 in those weaker, predominantly atomic components, suggestive of reduced shielding and/or stronger ambient radiation fields there.
In the stronger \ion{K}{1} components, where $f$(H$_2$) is higher, the implied $W$(5797.1)/$W$(5780.5) $\sim$ 1.1, which is comparable to the highest values measured for that ratio (e.g., Vos et al. 2011).
The $\lambda$5780.5 DIB is weak, relative to $N$(\ion{K}{1}), in the main M82 components, but is consistent with the mean Galactic trend in the weaker, higher velocity components.
The heavier relative weighting of the stronger \ion{K}{1} components (where CH is more abundant) for the $\lambda$4963.9 C$_2$-DIB, compared to the relative weighting for the $\lambda$5780.5 DIB, is consistent with the picture that the C$_2$-DIBs trace gas with higher molecular fractions, while the broader ''standard'' DIBs trace primarily atomic gas.
A similar situation may be present toward SN 2008fp (ESO428-G14), where the weaker than ''expected'' $\lambda$5780.5 DIB appears to be slightly offset from the strongest component seen in CN and \ion{K}{1} (Phillips et al. 2013; Cox \& Patat 2014).

\subsection{Predicting $N$(H), $E(B-V)$, and $A_{\rm V}$}
\label{sec-pred}

Given the relationships found between the DIBs and various other constituents of the ISM in our Galaxy and in the Magellanic Clouds, it is of interest to see whether observations of the DIBs can be used to estimate $N$(H), $E(B-V)$, and $A_{\rm V}$ when direct measurements of those quantities are not available or feasible.
For example, the generally good, nearly linear correlation between $W$(5780.5) and $N$(H) in the Galactic ISM ($r$ = 0.90; slope $\sim$ 1.05; Friedman et al. 2011; Welty 2014) suggests that reasonable estimates for $N$(H) might be obtained from measurements of the $\lambda$5780.5 DIB, both for sight lines where UV spectra of Lyman-$\alpha$ absorption are not available (e.g., toward SN 2014J, as in Sec.~\ref{sec-shift} above) and toward stars of spectral type later than about B3 (where stellar Lyman-$\alpha$ absorption becomes strong).
The estimated $N$(H) could be too small, however, for sight lines characterized by strong radiation fields (as in the Orion Trapezium region) or by significantly sub-solar metallicities (as in the LMC and SMC), where the DIBs can be significantly weaker (Welty et al. 2006).
While estimates for $E(B-V)$ and $A_{\rm V}$ based on the Galactic relationships with the DIB equivalent widths should not be as significantly affected by differences in metallicity, the ''standard'' DIBs can be weaker than expected -- and the reddening and extinction thus underestimated -- when the molecular fraction is large.
Examination of the DIB equivalent widths versus $N$(\ion{K}{1}) and/or $N$(\ion{Na}{1}) [and of the $N$(CN)/$N$(CH) ratio, when available] may aid in identifying such cases -- and may also enable some refinement of the estimated $E(B-V)$ and $A_{\rm V}$.

%%%%%  Table 4  %%%%%
\begin{deluxetable}{rccccccc}
\tablecolumns{8}
\tabletypesize{\scriptsize}
\tablecaption{Predicted $E(B-V)$ and $A_{\rm V}$ \label{tab:pred}}
\tablewidth{0pt}

\tablehead{
\multicolumn{1}{c}{DIB}&
\multicolumn{1}{c}{FWHM\tablenotemark{a}}&
\multicolumn{1}{c}{$W$(M82)\tablenotemark{b}}&
\multicolumn{1}{c}{$W$(M82)\tablenotemark{c}}&
\multicolumn{1}{c}{$E(B-V)$\tablenotemark{d}}&
\multicolumn{1}{c}{$E(B-V)$\tablenotemark{e}}&
\multicolumn{1}{c}{$A_{\rm V}$\tablenotemark{f}}&
\multicolumn{1}{c}{$A_{\rm V}$\tablenotemark{g}}\\
\multicolumn{1}{c}{}&
\multicolumn{1}{c}{(\AA)}&
\multicolumn{1}{c}{(meas)}&
\multicolumn{1}{c}{(pred)}&
\multicolumn{1}{c}{(pred)}&
\multicolumn{1}{c}{(corr)}&
\multicolumn{1}{c}{(pred)}&
\multicolumn{1}{c}{(corr)}\\
\multicolumn{1}{c}{(1)}&
\multicolumn{1}{c}{(2)}&
\multicolumn{1}{c}{(3)}&
\multicolumn{1}{c}{(4)}&
\multicolumn{1}{c}{(5)}&
\multicolumn{1}{c}{(6)}&
\multicolumn{1}{c}{(7)}&
\multicolumn{1}{c}{(8)}}

\startdata
%% DIB    FWHM      M82      pred   EBVp   EBVc    Avp    Avc
4963.9  & 0.62 &  23$\pm$3  &  22 & 0.74 & 0.77 & 2.23 & 2.18 \\
5487.7  & 5.20 &  89$\pm$13 &  75 & 0.80 & 0.60 & 2.08 & 1.79 \\
5705.1  & 2.58 &  55$\pm$8  &  71 & 0.49 & 0.52 & 1.24 & 1.54 \\
5780.5  & 2.11 & 319$\pm$18 & 346 & 0.62 & 0.67 & 1.81 & 1.95 \\
5797.1  & 0.77 & 210$\pm$6  & 151 & 1.18 & 0.87 & 3.22 & 2.64 \\
6196.0  & 0.42 &  46$\pm$5  &  40 & 0.92 & 0.76 & 2.42 & 2.24 \\
6203.6  & 4.87 & 150$\pm$14 & 130 & 0.83 & 0.65 & 2.12 & 1.88 \\
6283.8  & 4.77 & 866$\pm$48 & 906 & 0.75 & 0.69 & 2.06 & 2.09 \\
6379.3  & 0.58 &  41$\pm$4  &  73 & 0.43 & 0.69 & 1.31 & 1.80 \\
6613.6  & 0.93 & 226$\pm$5  & 176 & 1.07 & 0.85 & 2.70 & 2.56 \\
\hline
average &   &   &   & 0.78$\pm$0.24 & 0.71$\pm$0.11 & 2.12$\pm$0.59 & 2.07$\pm$0.35 \\
\enddata
\tablenotetext{a}{FWHM measured toward HD~204827 (Hobbs et al. 2009).}
\tablenotetext{b}{$W$(DIB) for M82 (from Table 1).}
\tablenotetext{c}{Predicted $W$(DIB) for M82, assuming log[$N$(\ion{K}{1})] = 12.28 and Galactic relationships given in Appendix Table~\ref{tab:corr}.}
\tablenotetext{d}{Predicted M82 $E(B-V)$ from M82 $W$(DIB) and Galactic relationships from Friedman et al. (2011).}
\tablenotetext{e}{Predicted M82 $E(B-V)$, adjusted for residuals [log $W$(meas) minus log $W$(pred)], using relationships given in Appendix Table~\ref{tab:corr}.}
\tablenotetext{f}{Predicted M82 $A_{\rm V}$ from M82 $W$(DIB) and relationships given in Appendix Table~\ref{tab:corr}.}
\tablenotetext{g}{Predicted M82 $A_{\rm V}$, adjusted for residuals [log $W$(meas) minus log $W$(pred)], using relationships given in Appendix Table~\ref{tab:corr}.}
\end{deluxetable}
%%%%%%%%%%

In light of the use of Type Ia SNe as ''standardizable candles'' for investigations of the expansion of the Universe and of the potential effects of intervening dust on the brightness and colors of those SNe, independent estimates for the host galaxy reddening and extinction would be valuable.
Phillips et al. (2013) compared host galaxy visual extinctions $A_{\rm V}$, derived from optical and near-IR photometry, with host galaxy $N$(\ion{Na}{1}) toward 32 extragalactic Type Ia SNe; they also determined $N$(\ion{K}{1}) and/or $W$(5780.5) for roughly one-third of the sample.
Phillips et al. found that where the $\lambda$5780.5 DIB is detected, predictions of $A_{\rm V}$ based on $W$(5780.5) are more consistent with the adopted host galaxy $A_{\rm V}$ than are the corresponding predictions from either the equivalent widths of the \ion{Na}{1} D-lines ($\lambda\lambda$5890.0, 5895.9) or $N$(\ion{Na}{1}).
Such comparisons may be affected by several issues, however:
(1) the equivalent widths of the strong \ion{Na}{1} D lines can severely underestimate the amount of \ion{Na}{1} present when the bulk of the material is concentrated in a small number of narrow, saturated components;
(2) fits to the profiles of strong \ion{Na}{1} D lines in moderate-resolution spectra can yield estimates for $N$(\ion{Na}{1}) that are either too high or too low, depending on the adopted component structure (particularly the adopted component $b$-values);
(3) the fiducial Galactic relationships used by Phillips et al. were biased by unrecognized saturation effects in the Galactic \ion{Na}{1} column densities (see footnote 2 above); and
(4) the $\lambda$5780.5 DIB traces primarily atomic gas, while $A_{\rm V}$ is due to dust associated with both atomic and molecular gas.
Where the \ion{Na}{1} D lines are saturated, observations of the weaker \ion{K}{1} $\lambda\lambda$7664.9, 7699.0 and \ion{Na}{1} $\lambda$3302 lines can help to constrain both the component structures and the overall $N$(\ion{Na}{1}).
While the extragalactic sample is still quite limited [particularly for measurements of $N$(\ion{K}{1}) and $W$(DIBs)], there are a number of cases where $N$(\ion{Na}{1}), $N$(\ion{K}{1}), $W$(5780.5), and $A_{\rm V}$ are reasonably consistent with the Galactic trends.
There are, however, some host galaxy sight lines -- particularly with log[$N$(\ion{Na}{1})] $\sim$ 13.0 and log[$N$(\ion{K}{1})] $\sim$ 11.0 -- where $A_{\rm V}$ is much lower than ''expected'' (Phillips et al. 2013), and some others -- typically with significant molecular content-- where $W$(5780.5) is low (Sec.~\ref{sec-m82} above).
Comparisons among the various tracers can help to identify such ''discrepant'' cases.

As the equivalent widths of other DIBs also appear to be reasonably well correlated with $E(B-V)$ and $A_{\rm V}$ (e.g., Friedman et al. 2011; Welty 2014; see also Appendix Table~\ref{tab:corr}), similar estimates for the reddening and extinction may be derived from measurements of those other DIBs.
Columns 5 and 7 of Table~\ref{tab:pred} list such estimates for the M82 $E(B-V)$ and $A_{\rm V}$ toward SN 2014J, derived from the equivalent widths of the ten DIBs considered in this paper.
As noted above, comparisons of the M82 DIB equivalent widths and $N$(\ion{K}{1}) with the Galactic trends for $W$(DIB) versus $N$(\ion{K}{1}) suggest that some DIBs ($\lambda\lambda$5705.1, 6379.3) will yield underestimates for $E(B-V)$ and $A_{\rm V}$, while several others ($\lambda\lambda$5797.1, 6613.6) will yield overestimates.
The values of $E(B-V)$ estimated from the ten DIBs range from 0.43 to 1.18; the mean $\pm$ standard deviation is 0.78$\pm$0.24.
Adjustment of the various estimated $E(B-V)$ for the residuals seen in $W$(DIB) versus $N$(\ion{K}{1}) narrows the range from 0.52 to 0.87, with mean $\pm$ standard deviation 0.71$\pm$0.11 (column 6 of Table~\ref{tab:pred}).
For $A_{\rm V}$, the initial predictions range from 1.24 to 3.22, with mean $\pm$ standard deviation 2.12$\pm$0.59; the adjusted values range from 1.54 to 2.64, with mean $\pm$ standard deviation 2.07$\pm$0.35 (column 8 of Table~\ref{tab:pred}).
Within the mutual uncertainties, the estimate for the host galaxy $E(B-V)$ is consistent with the values obtained from $N$(\ion{K}{1}) ($\sim$ 0.70) and $N$(\ion{Na}{1}) ($\sim$ 0.62), and with the upper limit ($<$ 0.8) estimated from optical and near-IR photometry of the SN by Polshaw et al. (2014), but is lower than the more recent values $\sim$ 1.2--1.3 derived by Goobar et al.(2014), Amanullah et al. (2014), and Foley et al. (2014).
The estimate for $A_{\rm V}$ is consistent with the value $\sim$ 2.1 obtained from $N$(\ion{K}{1}), and is slightly higher than, but consistent with the value $\sim$ 1.6 obtained from $N$(\ion{Na}{1}) and the values $\sim$ 1.7--2.0 inferred from the photometry (Goobar et al. 2014; Amanullah et al. 2014; Foley et al. 2014; Marion et al. 2014).
As our DIB-based estimates for $E(B-V)$ and $A_{\rm V}$ both employ the observed Galactic trends versus $W$(DIBs), the two values are not entirely independent, so that the corresponding ratio of total to selective extinction, $R_{\rm V}$ = $A_{\rm V}$/$E(B-V)$ $\sim$ 2.9, reflects the typical Galactic value $\sim$ 3.1.

As summarized, for example, by Phillips (2012), there are indications from fitting optical and near-IR photometry of Type Ia SNe that the host galaxy dust appears to be characterized by $R_{\rm V}$ consistent with Galactic values ($\sim$ 2.5--3.0) for relatively lightly reddened SNe, but by significantly smaller $R_{\rm V}$ ($\sim$ 1.5--2.0) for the most heavily reddened SNe (e.g., Folatelli et al. 2010; Mandel et al. 2012).
No such dependence of $R_{\rm V}$ on $A_{\rm V}$ is seen in our Galaxy, and there are no Galactic sight lines with R$_{\rm V}$ $<$ 2.0 in the surveys of Valencic et al. (2004) or Fitzpatrick \& Massa (2007).
Many of the sight lines in those surveys sample fairly long path lengths (and multiple interstellar clouds), however, so that such extreme cases might not be easily discernible; one of the Galactic sight lines with lowest $R_{\rm V}$ ($\sim$ 2.0--2.4), toward HD~210121, is dominated by a single cloud with fairly strong CN absorption and steep far-UV extinction (Welty \& Fowler 1992).
Goobar (2008) has suggested that the lower $R_{\rm V}$ inferred toward some SNe could be due to multiple scattering within a dusty circumstellar envelope.
The variable \ion{Na}{1} absorption seen toward the heavily reddened SN 2006X (with $R_{\rm V}$ $\sim$ 1.3--1.5), ascribed to changes in ionization in circumstellar material by Patat et al. (2007), has been taken to support the plausibility of that multiple scattering model.
In that case, however, the variations are for relatively weak higher-velocity \ion{Na}{1} components -- and not for the saturated main components (which presumably contain most of the dust).
Amanullah et al. (2014) find a similarly low $R_{\rm V}$ $\sim$ 1.3--1.4 for the host galaxy material toward SN 2014J; Foley et al. (2014) propose that the extinction and reddening are due to a combination of interstellar and circumstellar dust.
There have been no discernible variations in any of the absorption features seen in the optical spectra of SN 2014J that might be indicative of circumstellar material, however (Ritchey et al. 2014b; Foley et al. 2014), and upper limits on x-ray and radio emission have set corresponding fairly stringent limits on the density of material in the immediate vicinity of the SN (Margutti et al. 2014; P\'{e}rez-Torres et al. 2014).

\section{SUMMARY / CONCLUSIONS}
\label{sec-sum}

We have discussed the interstellar absorption features found in moderately high resolution, high S/N ratio optical spectra of SN 2014J (in the nearby galaxy M82), obtained with the ARC echelle spectrograph at Apache Point Observatory between 2014 January 27 and March 04 (bracketing the maximum V-band brightness of the SN).
Complex absorption from \ion{Na}{1}, \ion{K}{1}, \ion{Ca}{1}, and \ion{Ca}{2} is seen for LSR velocities between about $-$53 and +257 km~s$^{-1}$.
The absorption at $v_{\rm LSR}$ $\la$ 30 km~s$^{-1}$ is due to gas in the Galactic disk and halo; the absorption at higher velocities arises in gas associated with M82.
Absorption from CH, CH$^+$, and/or CN is seen for the strongest M82 components between 80 and 120 km~s$^{-1}$.
Many of the diffuse interstellar bands are also detected, at velocities corresponding to gas in M82.

Comparisons of the interstellar absorption in M82 with trends seen in the local Galactic ISM, in the lower metallicity Magellanic Clouds, and in other galaxies probed by SNe reveal both similarities and some intriguing differences. 
Overall, the $N$(\ion{K}{1})/$N$(\ion{Na}{1}) and $N$(CH)/$N$(\ion{K}{1}) ratios are very similar to those seen in our Galaxy; $N$(CN) is also quite consistent with the values seen locally, for comparable $N$(CH).
The $N$(\ion{Ca}{1})/$N$(\ion{K}{1}) ratio is high, suggestive of relatively mild depletion of calcium in M82, particularly in the higher velocity components.
The $N$(CH$^+$)/$N$(CH) ratio is very high -- significantly exceeded in the Galactic ISM only in several sight lines in the Pleiades.
The moderate $N$(CN)/$N$(CH) and very high $N$(CH$^+$)/$N$(CH) suggest that the molecular material toward SN 2014J is unlikely to be in very cold, dense clouds.
Of the ten DIBs considered in this paper, six ($\lambda\lambda$4963.9, 5487.7, 5780.5, 6196.0, 6203.6, 6283.8) have equivalent widths within $\pm$20\% of the mean Galactic values for the observed $N$(\ion{K}{1}); $\lambda$5705.1 and $\lambda$6379.3 are weaker by 20--45\%; and $\lambda$5797.1 and $\lambda$6613.6 are stronger by 25--40\%.
The overall $W$(5797.1)/$W$(5780.5) ratio is thus fairly high ($\sim$0.7) -- suggestive of relatively weak ambient radiation fields and/or shielded environments.
The combination of a high $N$(CH$^+$)/$N$(CH) ratio and a high $W$(5797.1)/$W$(5780.5) ratio is very unusual, as those two ratios exhibit a fairly tight anticorrelation in the Galactic ISM (and elsewhere).

While $W$(5780.5) is moderately correlated with both $N$(\ion{K}{1}) ($r$ $\sim$ 0.71) and $E(B-V)$ ($r$ $\sim$ 0.82) in the Galactic ISM, the DIB is weaker than expected, relative to $N$(\ion{K}{1}) and/or $E(B-V)$, in a number of Galactic and extragalactic sight lines; the residuals (with respect to the mean Galactic trends) for $W$(5780.5) versus $N$(\ion{K}{1}) and $W$(5780.5) versus $E(B-V)$ are correlated.
In general, the sight lines in which the $\lambda$5780.5 DIB appears to be weak also exhibit fairly high $N$(CN) and/or fairly high molecular fractions $f$(H$_2$).
Similar behavior is exhibited by the other ''standard'' DIBs considered in this paper -- particularly the broader DIBs.
Those DIBs appear weak, relative to $N$(\ion{K}{1} and/or $E(B-V)$, in sight lines with significant molecular fractions, at least in part because they trace primarily atomic gas, whereas \ion{K}{1} and $E(B-V)$ trace both atomic and molecular material.
The $\lambda$4963.9 ''C$_2$-DIB'' is not weaker than expected in those sight lines, however, because the C$_2$-DIBs can be present (and enhanced) in molecular gas.

The profiles of many of the M82 DIBs appear to be shifted in velocity, relative to the envelope of the \ion{K}{1} profiles toward SN 2014J, with different shifts for different DIBs.
The DIB profiles toward SN 2014J may be modeled by combining ''intrinsic'' DIB profiles (derived from ARCES spectra of 20 Aql) for all the velocity components seen in \ion{K}{1}.
Uniform weighting of the \ion{K}{1} components by $N$(\ion{K}{1}) yields a good match to the observed DIB profile only for the $\lambda$4963.9 C$_2$-DIB; heavier relative weighting of the weaker, higher velocity \ion{K}{1} components is required to fit the observed profiles of the other ''standard'' DIBs -- e.g., by factors of about 3 for the $\lambda$5797.1 DIB and about 7 for the $\lambda$5780.5 DIB.
The differences in relative weighting [i.e., in $W$(DIB)/$N$(\ion{K}{1})] are suggestive of differences in local physical/environmental conditions in the stronger ''main'' M82 components and in the weaker, higher velocity components -- and of differences in the responses of the various DIBs to those local conditions.
If standard Galactic relationships between $W$(5780.5) and $N$(H) and between $N$(CH) and $N$(H$_2$) are used to predict $N$(H) and $N$(H$_2$) in those two component groups, we estimate that the stronger main components would contain about 65\% of the total hydrogen in the sight line, with $f$(H$_2$) $\sim$ 0.75 and $W$(5797.1)/$W$(5780.5) $\sim$ 1.1 and the weaker, higher velocity components would contain about 70\% of the total atomic hydrogen, with $f$(H$_2$) $\sim$ 0.15 and $W$(5797.1)/$W$(5780.5) $\sim$ 0.45.
The $\lambda$4963.9 C$_2$-DIB and (to a lesser degree) the $\lambda$5797.1 DIB are stronger in the main, largely molecular components, while the $\lambda$5780.5 and $\lambda$6613.6 ''standard'' DIBs are stronger in the higher velocity, primarily atomic components.

The correlation between the residuals, relative to the mean Galactic trends, of $W$(5780.5) versus $N$(\ion{K}{1}) and $W$(5780.5) versus $E(B-V)$ suggests that comparisons of $W$(5780.5) with $N$(\ion{K}{1}) may be used to identify cases where estimates of $E(B-V)$ and $A_{\rm V}$ based on the measured $W$(5780.5) would be too small -- and then to refine those estimates for $E(B-V)$ and $A_{\rm V}$.
%Application of such a procedure yields better agreement between the host galaxy $E(B-V)$ estimated from $W$(5780.5) and the values derived from optical and near-IR photometry for several of the SNe in the sample of Phillips et al. (2013).
Estimates of the reddening due to dust in M82 along the sight line to SN 2014J, derived from the equivalent widths of the ten DIBs considered in this paper (and Galactic trends of reddening versus DIB strength), yield $E(B-V)$ $\sim$ 0.71$\pm$0.11 mag -- consistent with the values estimated from $N$(\ion{Na}{1}) and $N$(\ion{K}{1}) and also with the lower end of the values inferred from photometry of the SN.
Corresponding estimates of the visual extinction, $A_{\rm V}$ $\sim$ 1.9$\pm$0.2 (from the DIBs, \ion{Na}{1}, and \ion{K}{1}), are consistent with the most recent values inferred from the photometry.
The explanation for the low $R_{\rm V}$ inferred toward SN 2014J (and many other Type Ia SNe) remains elusive.

\acknowledgments

Support for this work has been provided by the National Science Foundation, under grants AST-1009603 (DGY), AST-1008424 (JAD), and AST-1238926 (DEW).
AMR gratefully acknowledges support from the Kenilworth Fund of the New York Community Trust.

{\it Facilities:} \facility{ARC (ARCES)}

\appendix

%%%%%%%%%%
\begin{deluxetable}{rrrrrrrrrrrrrr}
\tablecolumns{14}
\tabletypesize{\scriptsize}
\tablecaption{Correlation Fit Coefficients \label{tab:corr}}
\tablewidth{0pt}

\tablehead{
\multicolumn{1}{c}{DIB}&
\multicolumn{1}{c}{FWHM}&
\multicolumn{3}{c}{$E(B-V)$\tablenotemark{a}}&
\multicolumn{3}{c}{$A_{\rm V}$\tablenotemark{a}}&
\multicolumn{3}{c}{$N$(\ion{K}{1})\tablenotemark{a}}&
\multicolumn{3}{c}{Residual Plots\tablenotemark{b}}\\
\multicolumn{1}{c}{ }&
\multicolumn{1}{c}{(\AA)}&
\multicolumn{1}{c}{$r$}&
\multicolumn{1}{c}{a}&
\multicolumn{1}{c}{b}&
\multicolumn{1}{c}{$r$}&
\multicolumn{1}{c}{a}&
\multicolumn{1}{c}{b}&
\multicolumn{1}{c}{$r$}&
\multicolumn{1}{c}{a}&
\multicolumn{1}{c}{b}&
\multicolumn{1}{c}{$r$}&
\multicolumn{1}{c}{a}&
\multicolumn{1}{c}{b}\\
\multicolumn{1}{c}{(1)}&
\multicolumn{1}{c}{(2)}&
\multicolumn{1}{c}{(3)}&
\multicolumn{1}{c}{(4)}&
\multicolumn{1}{c}{(5)}&
\multicolumn{1}{c}{(6)}&
\multicolumn{1}{c}{(7)}&
\multicolumn{1}{c}{(8)}&
\multicolumn{1}{c}{(9)}&
\multicolumn{1}{c}{(10)}&
\multicolumn{1}{c}{(11)}&
\multicolumn{1}{c}{(12)}&
\multicolumn{1}{c}{(13)}&
\multicolumn{1}{c}{(14)}}

\startdata
%%                     EBV                     Av                       K I                      resid
%%       fwhm    r      a       b       r      a       b       r         a       b       r         a       b
4963.9 & 0.62 & 0.79 & 1.465 & 1.140 & 0.76 & 1.010 & 0.990 & 0.84 & $-$7.200 & 0.695 & 0.55 &    0.050 & 0.920 \\
5487.7 & 5.20 & 0.79 & 2.090 & 0.895 & 0.80 & 1.665 & 0.900 & 0.58 & $-$3.105 & 0.405 & 0.74 & $-$0.040 & 0.755 \\
5705.1 & 2.58 & 0.80 & 2.055 & 0.815 & 0.81 & 1.660 & 0.880 & 0.68 & $-$3.165 & 0.415 & 0.76 &    0.000 & 0.750 \\
5780.5 & 2.11 & 0.82 & 2.690 & 0.935 & 0.85 & 2.255 & 0.945 & 0.71 & $-$3.355 & 0.480 & 0.68 &    0.000 & 0.690 \\
5797.1 & 0.77 & 0.84 & 2.290 & 0.985 & 0.84 & 1.825 & 0.970 & 0.84 & $-$4.760 & 0.565 & 0.45 &    0.045 & 0.625 \\
6196.0 & 0.42 & 0.85 & 1.735 & 0.945 & 0.86 & 1.285 & 0.965 & 0.84 & $-$4.785 & 0.520 & 0.58 &    0.010 & 0.625 \\
6203.6 & 4.87 & 0.83 & 2.295 & 0.885 & 0.80 & 1.885 & 0.885 & 0.55 & $-$2.985 & 0.415 & 0.73 & $-$0.010 & 0.720 \\
6283.8 & 4.77 & 0.82 & 3.085 & 0.820 & 0.77 & 2.685 & 0.830 & 0.67 & $-$1.835 & 0.390 & 0.75 &    0.005 & 0.665 \\
6379.3 & 0.58 & 0.73 & 1.940 & 0.960 & 0.70 & 1.460 & 1.280 & 0.84 & $-$6.555 & 0.685 &  ... &   (0.00) &(0.70) \\
6613.6 & 0.93 & 0.83 & 2.380 & 1.195 & 0.87 & 1.845 & 1.095 & 0.85 & $-$6.535 & 0.715 & 0.49 &    0.030 & 0.545 \\
\enddata
\tablenotetext{a}{Relations are log[$W$(DIB)] = a + b $\times$ log[x], where x = $E(B-V)$, $A_{\rm V}$, or $N$(\ion{K}{1}).}
\tablenotetext{b}{Residuals are observed log[$W$(DIB)] minus value for mean Galactic relationship.
Fits are to residual with respect to log[$E(B-V)$] versus residual with respect to log[$N$(\ion{K}{1})].
The linear correlation coefficients ($r$) are also given for each relationship. 
Representative values for the slope and intercept were used for the $\lambda$6379.3 DIB residuals.}
\end{deluxetable}
%%%%%%%%%%

\section{Correlation Fits}
\label{sec-corr}

Friedman et al. (2011) investigated correlations between eight DIBs and $N$(H), $N$(H$_2$), and $E(B-V)$, for a sample of 133 Galactic sight lines observed with ARCES and listed coefficients found for linear fits to both log[$N$(H)] versus log[$W$(DIB)] and $E(B-V)$ versus $W$(DIB).
We have performed similar comparisons of log[$W$(DIB)] versus log[$E(B-V)$], log[$A_{\rm V}$], and log[$N$(\ion{K}{1})], using both the Friedman et al. sample [augmented by equivalent widths for the $\lambda$4963.9 and $\lambda$6379.3 DIBs from Thorburn et al. (2003)] and a somewhat larger sample of Galactic data now available. 
The values for the visual extinction $A_{\rm V}$ have been compiled by B. Rachford (2002, priv. comm.) for our database of DIB measurements and related interstellar quantities (e.g., Friedman et al. 2011); additional values were taken from Fitzpatrick \& Massa (2007) and Valencic et al. (2004).
For the (mostly lightly reddened) Galactic sight lines where $A_{\rm V}$ has not been independently determined, we have assumed $A_{\rm V}$ = 3.1 $\times$ $E(B-V)$.
The fit coefficients listed in columns 4--5, 7--8, and 10--11 of Table~\ref{tab:corr} are the means of the values obtained for weighted and unweighted fits to the data in the augmented Friedman et al. sample (e.g., Welty \& Hobbs 2001).
In general, the various ''discrepant'' sight lines (Sco-Oph, Trapezium, and all extragalactic) were not included in the fits; sight lines with $E(B-V)$ $<$ 0.05 mag and $A_{\rm V}$ $<$ 0.15 mag were excluded from the fits involving those quantities.
The notes to Tables~\ref{tab:dibs} and \ref{tab:pred} indicate which relationships were used to predict the Galactic and M82 DIB strengths from the respective $N$(\ion{K}{1}) and to predict the M82 $E(B-V)$ and $A_{\rm V}$ from the DIB equivalent widths.
The last three columns of Table~\ref{tab:corr} describe the correlations found between the residuals (observed minus mean) of the DIBs versus $E(B-V)$ and $N$(\ion{K}{1}), which were used to refine the estimates for $E(B-V)$.
As noted in the text, stronger correlations between the residuals (higher $r$ values) are found for the broader DIBs.
For similar sample sizes and dynamic ranges, the correlations between $E(B-V)$ and $A_{\rm V}$ and the various $W$(DIB) are characterized by similar values of $r$.
The lower $r$ values noted for $A_{\rm V}$ versus $W$(DIB) by Welty (2014) were for the more limited sample of independently determined $A_{\rm V}$.

\end{document}